\pdfoutput=1
\documentclass[journal,twocolumn]{IEEEtran}
\usepackage{amsmath}
\usepackage{algorithm,algpseudocode}
\usepackage{amssymb}
\usepackage{color}
\usepackage{enumerate}
\usepackage{wrapfig}
\usepackage{graphicx}          % Include this line if your 
\usepackage{cite}
\usepackage{hyperref}
\hypersetup{
	colorlinks = true,
	citecolor = cyan,
}
\usepackage{url}

\newcommand{\R}{\mathbb{R}}

\newcommand{\tu}[1]{\mathrm{#1}} 
\newcommand{\C}{\mathbb{C}}
\newcommand{\X}{\mathbb{X}}

\def\be{\begin{equation}}
	\def\ee{\end{equation}}
\def\ba{\begin{array}}
	\def\ea{\end{array}}
\newcommand{\smat}[1]{\left[\begin{smallmatrix}#1\end{smallmatrix}\right]}
\newcommand{\bmat}[1]{\begin{bmatrix}#1\end{bmatrix}}

\newcommand{\Obs}{\mathcal{O}}

\newcommand{\Ru}{\mathcal{R}_{\tu{u}}}
\newcommand{\Rw}{\mathcal{R}_{\tu{w}}}

\newtheorem{proposition}{Proposition}

\newtheorem{lemma}{Lemma}
\newtheorem{corollary}{Corollary}
\newtheorem{theorem}{Theorem}
\newtheorem{definition}{Definition}
\newtheorem{assumption}{Assumption}
\newtheorem{problem}{Problem}
\newtheorem{remark}{Remark}

\newenvironment{proof}{\begin{IEEEproof}}{\end{IEEEproof}}

\title{\LARGE \bf
	Data-Driven Output Regulation via Predictive Control
}
\author{Wenjie~Liu, Gang Wang,~\IEEEmembership{Senior Member,~IEEE}, and Lihua Xie,~\IEEEmembership{Fellow,~IEEE}
	\thanks{
		This work was supported by the Ministry of Education, Singapore, under AcRF Tier 1 Grant RG64/23, and the NTU-VinUni Joint Lab on Robotics and Embodied AI, and in part by the National Natural Science Foundation of China under Grants U23B2059 and 62688201. (\textit{Corresponding author:} Lihua Xie.)
	}
		\thanks{Wenjie Liu is with the School of Electrical and Electronic Engineering, Nanyang Technological University, 639798, Singapore (e-mail: wenjie.liu@ntu.edu.sg).}%
	\thanks{Gang Wang is with the State Key Lab of Autonomous Intelligent Unmanned Systems, School of Automation, Beijing Institute of Technology, Beijing 100081, China (e-mail: gangwang@bit.edu.cn).}%
	\thanks{Lihua Xie is with the NTU–VinUni Joint Research Laboratory for Robotics and Embodied AI, School of Electrical and Electronic Engineering, Nanyang Technological University, Singapore 639798 and VinUniversity, Hanoi 100000, Vietnam (e-mail: elhxie@ntu.edu.sg).}%
}

\begin{document}
	\maketitle
%\thispagestyle{empty}
%\pagestyle{empty}
% \linenumbers

\allowdisplaybreaks
	%\thispagestyle{empty}
	%\pagestyle{empty}

	%%%%%%%%%%%%%%%%%%%%%%%%%%%%%%%%%%%%%%%%%%%%%%%%%%%%%%%%%%%%%%%%%%%%%%%%%%%%%%%%
	\begin{abstract}
		This paper addresses constrained output regulation for unknown linear systems by means of predictive control directly from input-output data. The objective is to reject disturbances and track references generated by a known exosystem, without identifying a plant model or solving the output regulator equations. 
		The first step is a trajectory parametrization for plants driven by exosignals with known dynamics. % By exploiting the known exosystem eigenstructure, Willems et al's fundamental lemma is extended from autonomous input-output trajectories to trajectories coupled with exogenous modes.
		By exploiting the known exosystem eigenstructure, input-output trajectories driven by an exosignal of known phase are parametrized directly from data, which contains the affine version of Willems et al.'s fundamental lemma as a special case. This parametrization yields a nominal data-driven predictive control (DDPC) regulator, for which a Lyapunov argument proves exponential convergence to the regulator manifold. For periodic exosignals, the steady-state feedforward input is eliminated from the online problem through an incremental-input regularization, leading to an implementable regulator based on measured data and known exosystem information. 
		Bounded output noise is handled by slack and coefficient regularization, with a regularization center computed offline from data; the resulting robust DDPC law is recursively feasible and practically stabilizing, with an ultimate regulation error that vanishes with the noise bound. A constrained lateral vehicle-control example with periodic road curvature illustrates disturbance rejection, input-constraint satisfaction, and the transition from exact regulation with noise-free data to practical regulation with noisy data.
	\end{abstract}


\section{Introduction}\label{sec:intro}    

Model predictive control (MPC) is powerful because it combines performance
optimization with explicit constraint handling. Its success in process
control, robotics, reinforcement learning, and autonomous motion planning
has also motivated efforts to improve its computational efficiency
\cite{hou2024harnessing}. Its effectiveness, however, relies on the
availability of a predictive model that is accurate over the operating
region of interest \cite{rawlings2019model}. This modeling requirement
becomes a bottleneck for systems whose dynamics are high-dimensional,
strongly coupled, or repeatedly reconfigured. In such settings, deriving and
maintaining a first-principles model can dominate the control-design effort,
whereas input-output data are often readily available.

Data-driven predictive control (DDPC) has emerged as a principled way to reduce this modeling burden \cite{dorfler2022bridging,wang2026data}. A key enabler is \emph{Willems et al}'s fundamental lemma \cite{willems2005note}, which states that, under persistence of excitation, finite trajectories of an unknown linear system can be represented by linear combinations of previously measured trajectories. This observation has led to a set of MPC-like controllers with stability and robustness guarantees directly from data \cite{berberich2019data,berberich2021robust,Liu2023data,berberich2021linearII,chiuso2025harnessing,huang2023robust,xie2026data,alsalti2026robust}. 
Most existing DDPC results, nevertheless, are formulated around stabilization to a fixed equilibrium or tracking of equilibrium-like setpoints. Output regulation is different: the target behavior is an exosystem-driven trajectory on a regulator manifold, and the corresponding steady-state input is characterized by regulator equations that depend on the unknown plant.

Output regulation is the natural language for many constrained control tasks: grid converters must reject periodic load variations \cite{liu2025survey}, vehicles must follow curvature-varying roads, and networked or robotic systems often have to synchronize with time-varying leaders. 
Classical solutions rely on the internal model principle and on solvability of the output regulator equations \cite{wonham1974linear,francis1976internal,davison1976robust,huang2004nonlinear}. These results are foundational, but their implementation typically requires plant matrices and feedforward regulator gains. When the plant is unknown or only measured through noisy input-output data, identifying a model and then solving the output regulator equations introduces an additional approximation layer precisely in the part of the design that determines the steady-state behavior.

Recent data-driven approaches to output regulation are compared in Table \ref{tab:comparison}. Most of them are feedback-synthesis results and do not handle input constraints \cite{Jiao2021ddor,Zhu2024output-regulation,li2024data,liu2025outreg,bisoffi2026output}, whereas DDPC formulations for repetitive tasks \cite{dinkla2024data} lack closed-loop guarantees. Developing a constrained, data-driven predictive regulator with guarantees requires overcoming three obstacles.
First, the Hankel
prediction model must encode both the unknown plant response and the
temporal structure induced by the known exosystem eigenstructure.
{Second}, even with such a predictor, the usual equilibrium-centered DDPC stage
cost is not aligned with the regulator manifold, since maintaining this
manifold generally requires a nonzero feedforward input determined by
the unknown regulator equations. 
Third, with noisy data, the coefficient regularization must be centered at the nonzero steady-state coefficient rather than at zero.
These issues are addressed in the
present paper.
\color{black}

\begin{table}[]
	\centering
	\caption{Comparison with related output regulation approaches.}
	\label{tab:comparison}
	\setlength{\tabcolsep}{1.2pt}
	\begin{tabular}{lccccc}
		\hline
		& Data- & No regulator & Predictive & Guarantees with & Exosignal\\
		& driven & equations & control & noisy data & information\\
		\hline
		MPC \cite{kohler2021mpc-output-reg} & $\times$ & $\checkmark$ & $\checkmark$ & -- & $S$\\
		\cite{Jiao2021ddor,Zhu2024output-regulation} & $\checkmark$ & $\times$ & $\times$ & $\times$ & $S$\\
		\cite{li2024data} & $\checkmark$ & $\checkmark$ & $\times$ & $\checkmark$ & $S$\\
		\cite{liu2025outreg,bisoffi2026output} & $\checkmark$ & $\checkmark$ & $\times$ & $\checkmark$ & $S$\\
		DeePRC \cite{dinkla2024data} & $\checkmark$ & $\checkmark$ & $\checkmark$ & $\times$ & $T$\\
		This paper & $\checkmark$ & $\checkmark$ & $\checkmark$ & $\checkmark$ & $S$ and phase\\
		\hline
	\end{tabular}
\end{table}

We propose a predictive-control framework for data-driven output regulation of unknown linear systems driven by exogenous signals. The controller uses measured input-output trajectories, the exosystem dynamics, and online measurements; it does not identify the plant matrices and does not explicitly solve the classical output regulator equations. The analysis relies on the solvability of the output regulator equations, controllability/observability of the plant, a nonresonance condition, and incremental input-output-to-state stability of an auxiliary input-output realization. Under these standard structural conditions, the proposed receding-horizon laws steer the closed loop to the regulator manifold; under bounded output noise, they steer it to a neighborhood whose size is quantified by the noise level.

The main contributions are summarized as follows.
\begin{itemize}
%	\item[c1)] We extend Willems et al.'s fundamental lemma to input-output trajectories generated by unknown linear plants subject to exosignals with known dynamics. The construction uses a matrix factorization induced by the known
%	exosystem and produces a data-based prediction model that is consistent
%	with the temporal basis induced by the exosystem eigenstructure.
%	\item[c2)] We develop nominal DDPC formulations that use this parametrization inside receding-horizon constraints. One formulation mirrors model-based output-regulating MPC by regularizing the input around the steady-state feedforward action, which is not available; for periodic exosignals, a second formulation removes this unavailable quantity by using an incremental-input penalty.
%	\item[c3)] We establish recursive feasibility and exponential convergence to the regulator manifold for the nominal schemes. We also derive a robust counterpart with slack and coefficient regularization and prove practical stability under bounded output noise.
\item[c1)] A data-based parametrization of trajectories driven by exosignals
with known dynamics and known phase, built on the decomposition in
\cite{bisoffi2026output}; it recovers the affine fundamental lemma
\cite{berberich2021linearII} for $S=I$ and comes with an excitation
condition verifiable from the input and the exosystem basis.
\item[c2)] A constrained DDPC regulator that requires neither a plant model
nor the regulator equations; for periodic exosignals, the steady-state
feedforward input is removed by an incremental-input penalty.
\item[c3)] Recursive feasibility and exponential stability of the regulator
manifold with noise-free data, and practical stability under noisy data
and measurements, enabled by a regularization center computed offline from
data with a guaranteed mismatch bound.
%	\item[c4)] We validate the method on a constrained lateral vehicle-control problem with periodic road curvature, showing disturbance rejection, input-constraint satisfaction, and the predicted transition from exact to practical regulation when noisy data are used.
\end{itemize}

\emph{Notation.} We denote the sets of real and complex numbers by $\mathbb{R}$ and $\mathbb{C}$, respectively, and the sets of non-negative and positive integers by $\mathbb{N}$ and $\mathbb{N}_{+}$.
For any matrix $M$, $M^\dagger$ denotes its Moore--Penrose pseudoinverse. 
The notation $M \succ(\succeq)\, 0$ indicates that $M$ is positive (semi-) definite, and similarly, $M \prec (\preceq)\, 0$ denotes that $M$ is negative (semi-) definite. For vector $x \in \mathbb{R}^{n_x}$, $\| x\|$ denotes its Euclidean norm, and for a matrix $M$, $\| M \|$ denotes the spectral norm. The minimum and maximum eigenvalues of matrix $P$ are denoted by $\underline{\lambda}_P$ and $\overline{\lambda}_P$, respectively. 
% For $A_1 \in \C^{n_1 \times n_1}$, \dots, $A_k \in \C^{n_k \times n_k}$, the block diagonal matrix $\smat{A_1 & \dots & 0\\ \vdots & \ddots & \vdots\\ 0 & \dots & A_k}$ is written $A_1 \oplus \dots \oplus A_k$, which is the direct sum of matrices $A_1$, \dots, $A_k$.
% Additionally, for matrices $A$, $B$ and $C$ with compatible dimensions, we use the shorthand notation $AB \cdot C[\star]'$ to denote $ABC(AB)'$.
% For matrices $A = A'$, B, $C = C'$, we also abbreviate the symmetric matrix $\left[
% \begin{smallmatrix}
	% 	A & B\\
	% 	B' & C
	% \end{smallmatrix}\right]$ as $\left[
% \begin{smallmatrix}
	% A & B\\
	% \star & C
	% \end{smallmatrix}\right]$ or $\left[
% \begin{smallmatrix}
	% A & \star\\
	% B' & C
	% \end{smallmatrix}\right]$.
% The imaginary unit is $\i$.
% For $a \in \R$ and $b \in \R$, the complex conjugate of $a+ \i b$ is $\overline{a+ \i b} = a - \i b$.
% For integers $p$ and $q$ such that $p \ge q \ge 0$, ${p \choose q} := \frac{p!}{q!(p-q)!}$;
% for integers $p$ and $q$ such that $p \ge q$, $p \ge 0$ and $q < 0$, ${p \choose q} :=0$.
For a signal $x: \mathbb{N} \to \mathbb{R}^{n_x}$ and $T \in \mathbb{N}_+$, the vectorized form of the signal $x$ over the interval $[0, T-1]$ is denoted by $x_{[0,T-1]} := [x(0)'~x(1)'~\cdots~x(T-1)']'$.
The block-diagonal matrix with blocks $A_1, \ldots, A_N$ on its main diagonal is denoted ${\rm block.diag}(A_1, \ldots, A_N)$.
Given $N,L \in \mathbb{N}_+$, the Hankel matrix of order $L$ of a length-$N$ trajectory $x_{[0,N -1]}$ is denoted by $H_{L}(x_{[0,N -1]}):=\smat{
	x_{0} & x_{1} & \ldots & x_{N-L} \\
	\vdots & \vdots & \ddots & \vdots \\
	x_{L-1} & x_{L} & \ldots & x_{N-1}
}$.

% \section{Fundamental Lemma for systems with exosignal}

% In this section, we provide a data-driven parametrization of unknown systems subject to exosignal, i.e.,
% \begin{subequations}\label{eq:sys}
	% 	\begin{align}
		%     w_{t + 1} & = S w_t\label{eq:sys:w}\\
		% x_{t + 1} & = Ax_t + Bu_t + Ew_t \label{eq:sys:x}
		% \\
		%  y_t & = C x_t + F w_t\label{eq:sys:e}
		% \end{align}
	%  \end{subequations}
% where $x_t \in \mathbb{R}^{n}$ is the state, $u_t \in \mathbb{R}^{m}$ is the control input, $y_t \in \mathbb{R}^{p}$ is the output, and $w_t \in \mathbb{R}^{n_w}$ is the exosignal (e.g., external disturbance, and reference input).
% Matrices $A$, $B$, $C$, $E$, and $F$ are unknown constant matrices, while matrix $S$ is known with all eigenvalues lies on or outside the unit circle.

\section{Fundamental Lemma for LTI Systems with Exosignals}
\label{sec:fundamental-lemma}

This section establishes the data-based trajectory representation used throughout the paper. Consider the unknown linear system driven by an exosignal
\begin{subequations}\label{eq:sys}
	\begin{align}
		w_{t + 1} & = S w_t\label{eq:sys:w}\\
		x_{t + 1} & = Ax_t + Bu_t + Ew_t \label{eq:sys:x}
		\\
		y_t & = C x_t + F w_t\label{eq:sys:e}
	\end{align}
\end{subequations}
where $x_t \in \mathbb{R}^{n}$ is the state, $u_t \in \mathbb{R}^{m}$ is the control input, $y_t \in \mathbb{R}^{p}$ is the measured output, and $w_t \in \mathbb{R}^{n_w}$ represents disturbances or references generated by the exosystem.
The plant matrices $A$, $B$, $C$, $E$, $F$ and the particular trajectory $w_t$ are unknown, whereas the exosystem matrix $S$ and one input-output trajectory $\{u_t,y_t\}_{t = 0}^{N - 1}$ are available.
If $w_t=0$, Willems et al's fundamental lemma \cite{willems2005note} states that, under controllability and persistence of excitation, a candidate sequence $\{\bar{u}_t,\bar{y}_t\}_{t=0}^{L-1}$ is a valid trajectory if and only if there exists $g\in\mathbb{R}^{N-L+1}$ satisfying
\begin{equation}\label{eq:fund_lem_ori}
	\setlength{\abovedisplayskip}{3pt}
	\setlength{\belowdisplayskip}{3pt} 
	\smat{
		H_{L}\left(u_{[0,N - 1]}\right) \\
		H_{L}\left(y_{[0,N - 1]}\right)
	} g=\smat{
		\bar{u} \\
		\bar{y}
	}.
\end{equation}
For constant offsets, i.e., $S=I$, \cite[Theorem 1]{berberich2021linearII} accounts for the exogenous component by imposing the affine constraint $\sum_{i=0}^{N-L}g_i=1$.
Dynamic exosignals require a different treatment: the contribution of $w_t$ cannot in general be separated from the measured input-output data, but its temporal structure is known through $S$.
The following decomposition converts exosystem trajectories into a finite set of known basis signals with unknown coefficients.
\begin{lemma}[{\cite[Lemmas 3.3--3.4]{bisoffi2026output}}]
	\label{lem:forma-nuova-risposta-esosistema}
	Consider any exosignal sequence $\{w_t\}_{t=0}^{N-1}$ generated by
	\eqref{eq:sys:w} with initial condition $w_0\in\mathbb{R}^{n_w}$.
	Then, for $H_1(w)
	:= \bmat{
		w_0 & w_1 & \cdots & w_{N-L}
	}$, there exist an integer $n_M\le n_w$, a matrix
	$\mathcal{L}(w_0)\in\mathbb{R}^{n_w\times n_M}$ and known vectors
	$\mathcal{M}(i)\in\mathbb{R}^{n_M}$, $i=0,\ldots,N-L$, such that
	\begin{align}
		H_1(w)
		\!=\! \mathcal{L}(w_0)
		\smat{\!
			\mathcal{M}(0) \!& \mathcal{M}(1) & \cdots & \mathcal{M}(N\!-\!L)\!
		}
		\!:= \!\mathcal{L}(w_0)H_1(\mathcal{M})
		\label{eq:H1w=lH1M}
	\end{align}
	where $H_1(\mathcal{M})\in\mathbb{R}^{n_M\times (N-L+1)}$ is known and of
	full row rank.
\end{lemma}

\begin{remark}
	In \eqref{eq:H1w=lH1M}, the matrix $\mathcal{L}(w_0)$ depends on the
	initial condition $w_0$, whereas $H_1(\mathcal{M})$ is determined solely by
	the eigenstructure of the exosystem matrix $S$. In particular, the rows of
	$H_1(\mathcal{M})$ form an independent set of exosystem basis signals. The
	explicit construction of $H_1(\mathcal{M})$ follows from
	\cite[Claim C.1]{bisoffi2026output} and is omitted here for brevity.
	For instance, if $S$ is diagonalizable with all eigenvalues on the unit circle, as in the periodic setting of Section~\ref{sec:ddmpc:incre-mpc}, one can take $\mathcal M(t)$ as the stack of $\lambda_i^t$ for each distinct real eigenvalue $\lambda_i$ and of $(\cos(\theta_jt),\sin(\theta_jt))$ for each distinct complex-conjugate pair $e^{\pm{\rm i}\theta_j}$; e.g., for the exosystem in Section~\ref{sec:numerical_example}, $\mathcal M(t)=[1~\cos(\omega t)~\sin(\omega t)~\cos(2\omega t)~\sin(2\omega t)~\cos(3\omega t)~\sin(3\omega t)]^\top$. In general, Jordan blocks of $S$ additionally generate the terms $t^k\lambda^{t-k}$ \cite[Claim C.1]{bisoffi2026output}.
\end{remark}
% According to Lemma \ref{lem:forma-nuova-risposta-esosistema}, any exosignal sequence $\{w_t\}_{t = 0}^{N - 1}$ admits the representation
% \begin{align*}
	%    H_{1}(w) & =   \left[
	%     \begin{matrix}
		%         w_0 & w_1 & \ldots & w_{N - L}
		%     \end{matrix}
	%     \right] \\
	%     & =  \smat{\tilde{\mathcal{T}}L(w_0)M(0) & \tilde{\mathcal{T}}L(w_0)M(1) & \ldots & \tilde{\mathcal{T}}L(w_0)M(N - L)}\\ 
	%     & = \tilde{\mathcal{T}}L(w_0)\underbrace{\left[
		%     \begin{smallmatrix}
			%         M(0) & M(1) & \ldots & M(N - L)
			%     \end{smallmatrix}
		%     \right]}_{:= H_{1}(M)}
	% \end{align*}
% where matrix $M(i), \forall i = 0,\cdots, N - L$ is constructed from Lemma \ref{lem:forma-nuova-risposta-esosistema}.
% As in \cite{bisoffi2026output}, the matrices \( H_{1}(M) \) and $M(t)$ can contain redundant rows that increase the size of the optimization problem without adding predictive information. 
% We therefore retain only an independent set of exosystem basis rows.
% Specifically, let $H_1(\mathcal{M}) = \bmat{\mathcal{M}(0)\cdots\mathcal{M}(N - L)} \in \mathbb{R}^{n_M \times (N - L + 1)}$ and $\bar{M}_t \in \R^{n_M}$ be obtained from $H_1(M)$ and $M(t)$ in Lemma \ref{lem:forma-nuova-risposta-esosistema} by retaining, for every distinct real eigenvalue $\lambda_i$ and complex conjugate pair $(\rho_{r+i}e^{j\theta_{r+i}},\rho_{r+i}e^{-j\theta_{r+i}})$, the block rows with largest Jordan indices.
% Mathematically, there exists transfer matrix $\mathcal{T}_M \in \R^{n_M \times n_w}$ such that  $H_1(\mathcal{M}) = \mathcal{T}_M H_1(M)$ and $\bar{M}_t = \mathcal{T}_M M(t)$.

The data-based characterization also requires excitation not only of the control channel but also of the exosystem basis. This motivates the strengthened persistence-of-excitation condition.
% Next, we propose the following definition of persistence of excitation, where we write $\{ x_t\}_{t = 0}^{N - 1}$ for the state sequence corresponding to the available input-output data $\{u_t,y_t\}_{t = 0}^{N - 1}$.
\begin{definition}\label{def:pe}
	Consider system \eqref{eq:sys}.
	Let $\{ x_t\}_{t = 0}^{N - 1}$ be the state sequence corresponding to the available input-output data $\{u_t,y_t\}_{t = 0}^{N - 1}$.
	We say that the data $\{u_t,y_t \}_{t = 0}^{N - 1}$ are persistently exciting of order $L$ if 
	\begin{equation}\label{eq:def-pe-rank}
		{\rm rank} \left(
		\smat{
			H_1(x)\\
			H_{L}(u)\\
			H_1( \mathcal{M})
		}
		\right) = mL + n + n_M.
	\end{equation}
\end{definition}
% This definition is stronger than that in \cite{berberich2021linearII}, since we need to account for the exosignal \eqref{eq:sys:w} with $S \ne 0$.

Since the state sequence $x$ is not measured, the rank condition in
\eqref{eq:def-pe-rank} cannot be checked directly from the available
input-output data. To address this issue, the following proposition provides
a verifiable sufficient condition involving only the applied input and the
known exosystem basis.
\begin{proposition}
	\label{prop:pe-sufficient}
	Suppose that $(A,B)$ is controllable and $N\ge L+n$. Let
	$\mathcal M(\cdot)$ be the known exosystem basis associated with
	$\{w_t\}_{t=0}^{N-1}$ as specified in
	Lemma~\ref{lem:forma-nuova-risposta-esosistema}, and define
	$H_1(\mathcal M_{[0,N-L-n]})
	:=
	\smat{
		\mathcal M(0)&\mathcal M(1)&\cdots&\mathcal M(N-L-n)}$.
	If
	\begin{equation}\label{eq:relative-pe}
		{\rm rank}
		\smat{
			H_{L+n}(u)\\
			H_1(\mathcal M_{[0,N-L-n]})
		}
		=
		m(L+n)+n_M,
	\end{equation}
	then the data are persistently exciting of order $L$ in the sense of
	Definition~\ref{def:pe}, i.e., \eqref{eq:def-pe-rank} holds.
\end{proposition}
% \begin{proof}
%	See Appendix \ref{app:proof-prop1}.
%	See Appendix A in the Supplementary File.
% \end{proof}

\begin{proof}
	Let $\lambda\in\mathbb R^{1\times n}$,
	$\eta=\begin{bmatrix}\eta_0&\cdots&\eta_{L-1}\end{bmatrix}
	\in\mathbb R^{1\times mL}$, with
	$\eta_k\in\mathbb R^{1\times m}$, and
	$\rho\in\mathbb R^{1\times n_M}$ satisfy
	{\setlength{\abovedisplayskip}{3pt}
	\setlength{\belowdisplayskip}{3pt}
	\begin{equation}
		\label{eq:prop1:leftkernel}
		\begin{bmatrix}
			\lambda & \eta & \rho
		\end{bmatrix}
		\smat{
			H_1(x)\\
			H_L(u)\\
			H_1(\mathcal M)}
		=0.
	\end{equation}}
	We prove that $\lambda=0$, $\eta=0$, and $\rho=0$.
	
	By Lemma~\ref{lem:forma-nuova-risposta-esosistema}, we have $w_i=\mathcal L(w_0)\mathcal M(i)$.
	Moreover, the explicit construction of $\mathcal M(i)$ in
	\cite[Claim C.1]{bisoffi2026output} implies that the retained basis
	signals $\mathcal{M}(i)$ are invariant under the forward-shift
	operator. 
	In other words, there exists a known matrix
	$S_M\in\mathbb R^{n_M\times n_M}$, determined by $S$, such that $\mathcal M(i+j)=S_M^j\mathcal M(i), i,j\ge 0$.
	Set $D_M:=E\mathcal L(w_0)$. The state dynamics in \eqref{eq:sys:x} can then be written as $x_{i+1}=Ax_i+Bu_i+D_M\mathcal M(i)$.
	
	For $i=0,\ldots,N-L-n$, \eqref{eq:prop1:leftkernel} gives $\lambda x_i+\sum_{k=0}^{L-1}\eta_k u_{i+k}
	+\rho\mathcal M(i)=0$.
	Applying this relation at time $i+j$ and recursively substituting \eqref{eq:sys:x}, for $j=0,\ldots,n$, yields
	\begin{align}
		0={}&\lambda A^j x_i
		+\sum_{r=0}^{j-1}\lambda A^{j-1-r}Bu_{i+r}
		+\sum_{k=0}^{L-1}\eta_k u_{i+j+k} \notag\\
		&+\Big(
		\rho S_M^j+
		\sum_{r=0}^{j-1}\lambda A^{j-1-r}D_MS_M^r
		\Big)\mathcal M(i),
		\label{eq:left-kernel-shifted}
	\end{align}
	where an empty sum is understood as zero.
	
	Let $p_A(z)=\sum_{j=0}^{n}\beta_jz^j$ with $\beta_n=1$ be the characteristic polynomial of $A$. By the Cayley--Hamilton theorem,
	$\sum_{j=0}^{n}\beta_jA^j=0$. Multiplying
	\eqref{eq:left-kernel-shifted} by $\beta_j$ and summing over
	$j=0,\ldots,n$ therefore eliminates the state term and gives $\sum_{q=0}^{L+n-1}\bar\eta_q u_{i+q}
	+\bar\rho\mathcal M(i)=0$, 
	where $\bar\eta_q:=
	\sum_{\substack{0\le j\le n\\0\le q-j\le L-1}}
	\beta_j\eta_{q-j}
	+
	\sum_{j=q+1}^{n}\beta_j\lambda A^{j-1-q}B$
	and $\bar\rho:=
	\sum_{j=0}^{n}\beta_j
	\left(
	\rho S_M^j+
	\sum_{r=0}^{j-1}\lambda A^{j-1-r}D_MS_M^r
	\right)$.
	Thus, $\smat{\bar\eta_0&\cdots&\bar\eta_{L+n-1}&\bar\rho}
	\smat{
		H_{L+n}(u)\\
		H_1(\mathcal M_{[0,N-L-n]})
	}=0$.
	
	By \eqref{eq:relative-pe}, all the coefficient blocks above are zero.
	For $s=L-1,\ldots,0$, the equation
	$\bar\eta_{n+s}=0$ has leading term $\beta_n\eta_s=\eta_s$, while its
	remaining terms involve only $\eta_{s+1},\ldots,\eta_{L-1}$. Backward
	substitution therefore gives $\eta_0=\cdots=\eta_{L-1}=0$.
	
	With $\eta=0$, the equations
	$\bar\eta_{n-1}=0,\ldots,\bar\eta_0=0$ successively imply $\lambda B=\lambda AB=\cdots=\lambda A^{n-1}B=0$.
	Controllability of $(A,B)$ then yields $\lambda=0$.
	Returning to \eqref{eq:prop1:leftkernel} gives $\rho H_1(\mathcal M)=0$.
	Consequently, since $H_1(\mathcal M)$ has full row rank, $\rho=0$. 
	Therefore, the left kernel
	of the matrix in \eqref{eq:def-pe-rank} is trivial, which completes the proof.
\end{proof}

	Note that Proposition~\ref{prop:pe-sufficient} does not follow from applying the fundamental lemma to the extended input $(u,\mathcal M)$. Since $\mathcal M(\cdot)$ is generated autonomously, ${\rm rank}\,H_{L+n}(\mathcal M)\le n_M$, so the extended input can never be persistently exciting of order $L+n$. The proof above instead exploits the shift invariance $\mathcal M(i+j)=S_M^j\mathcal M(i)$ together with the Cayley--Hamilton theorem, and only requires excitation of the first block row $H_1(\mathcal M)$.
%\end{remark}

\color{black}

We can now state the trajectory parametrization, which plays an important role in the controller design later.
\begin{theorem}\label{thm:fund-lem-exo}
	Suppose the input-output data $u_{[0,N-1]}$, $y_{[0,N-1]}$ are generated by
	\eqref{eq:sys} and are persistently exciting of order $L$ in the sense of
	Definition \ref{def:pe}. Let $\bar M_t:=\mathcal M(t)$. Then a candidate
	sequence $\bar{u}_{[0,L-1]|t}, \bar{y}_{[0,L-1]|t}$ is an input-output
	trajectory of \eqref{eq:sys} corresponding to the exosystem basis vector
	$\bar M_t$, i.e., driven by the exosignal $\bar w_{k|t}=\mathcal L(w_0)\mathcal M(t+k)=w_{t+k}$, $k=0,\ldots,L-1$, of the same exosystem realization as the data at time index $t$, if and only if there exists $g_t\in\mathbb R^{N-L+1}$ such that
	\begin{equation}\label{eq:Hg=u:M-no-redundent}
		\begin{bmatrix}
			\begin{smallmatrix}
				H_{L}(y)\\
				H_{L}(u)\\
				H_1(\mathcal{M})
			\end{smallmatrix}
		\end{bmatrix} g_t = \begin{bmatrix}
			\begin{smallmatrix}
				\bar{y}_{[0,L - 1]|t}\\
				\bar{u}_{[0,L - 1]|t}\\
				\bar{M}_t
			\end{smallmatrix}
		\end{bmatrix}.
	\end{equation} 
\end{theorem}
\begin{proof}
	\textbf{Proof of ``if'':}
	For input-output data $\{u_t,y_t \}_{t = 0}^{N - 1}$ generated by \eqref{eq:sys}, it holds that 
	\begin{align}\label{eq:y-sequence-xuw-abbre}
		\setlength{\abovedisplayskip}{3pt}
		\setlength{\belowdisplayskip}{3pt} 
		y_{[i,i + L  - 1]} = \mathcal{O}_{L} x_i + \mathcal{T}^u_{L} u_{[i, i + L  - 1]} + \hat{\mathcal{T}}^w_{L}  \mathcal{L}(w_0) \mathcal{M}(i)
	\end{align}
	where
	\begin{align}
		\label{eq:matrix_O_T}
		&\mathcal{O}_L = \begin{bmatrix}
			\begin{smallmatrix}
				C\\
				CA\\
				CA^2\\
				\vdots\\
				CA^{L - 1}
			\end{smallmatrix}
		\end{bmatrix},~\mathcal{T}^{u}_{L} = \begin{bmatrix}
			\begin{smallmatrix}
				0 & 0 & \cdots & 0 & 0\\
				CB & 0 & \cdots & 0 & 0\\
				CAB & CB & \cdots & 0 & 0\\
				\vdots & \vdots & \ddots & \vdots & \vdots\\
				CA^{L - 2} B & CA^{L  - 3}B & \cdots & CB & 0
			\end{smallmatrix}
		\end{bmatrix}, \nonumber\\
		&\mathcal{\hat{T}}_{L}^w =\underbrace{\begin{bmatrix}
				\begin{smallmatrix}
					F & 0 & \cdots & 0 & 0\\
					CE & F & \cdots & 0 & 0\\
					CAE & CE & \cdots & 0 & 0\\
					\vdots & \vdots & \ddots & \vdots & \vdots\\
					CA^{L - 2} E & CA^{L - 3}E & \cdots & CE & F
				\end{smallmatrix}
		\end{bmatrix}}_{:= \mathcal{T}^w_{L}}
		\smat{I\\
			S\\
			S^2\\
			\vdots\\
			S^{L - 1}}
	\end{align}
	Indeed, \eqref{eq:H1w=lH1M} gives $w_i=\mathcal L(w_0)\mathcal M(i)$.
	Hence, if \eqref{eq:Hg=u:M-no-redundent} holds, we have that
	\begin{align*}
		&\bar{y}_{[0,L - 1]|t} \overset{\eqref{eq:Hg=u:M-no-redundent}}{=} \sum_{i = 0}^{N - L} y_{[i,i + L - 1]} g_{t,i}   \\
		& \overset{\eqref{eq:y-sequence-xuw-abbre}}{=} \sum_{i = 0}^{N - L} g_{t,i} (\mathcal{O}_{L} x_i \!+ \mathcal{T}^u_{L} u_{[i, i + L - 1]} \! + \hat{\mathcal{T}}^w_{L}  \mathcal{L}(w_0)  \mathcal{M}(i))\\
		& \overset{\eqref{eq:Hg=u:M-no-redundent}}{=} \mathcal{O}_{L} \sum_{i = 0}^{N - L} g_{t,i} x_i + \! \mathcal{T}^u_{L} \bar{u}_{[0,L  - 1]|t} +\! \hat{\mathcal{T}}^w_{L}  \mathcal{L}(w_0)  \bar{M}_t.
	\end{align*}
	This is the output response of \eqref{eq:sys} over the horizon
	$[0,L-1]$ under the input $\bar u_{[0,L-1]|t}$, with initial state $\bar x_{0|t}:=\sum_{i=0}^{N-L}g_{t,i}x_i$
	and exosignal initial value $\bar w_{0|t}:=\mathcal L(w_0)\bar M_t$.
	Hence, $\{\bar u_k,\bar y_k\}_{k=0}^{L-1}$ is an input-output trajectory of
	\eqref{eq:sys} corresponding to the exosystem basis vector $\bar M_t$.
	
	\textbf{Proof of ``only if'':}
	Let $\bar u_{[0,L-1]|t},\bar y_{[0,L-1]|t}$ be an input-output trajectory of
	\eqref{eq:sys} with initial state $x_t$ and exosignal basis vector
	$\bar M_t=\mathcal M(t)$. Since the data are persistently exciting of order
	$L$, the matrix $\smat{
		H_1(x)\\
		H_L(u)\\
		H_1(\mathcal M)}$
	has full row rank. Therefore, there exists $g_t\in\mathbb R^{N-L+1}$ such that
	\begin{equation}\label{eq:Hg=u:M-thm-pr}
		\begin{bmatrix}
			\begin{smallmatrix}
				H_{1}(x)\\
				H_{L}(u)\\
				H_1(\mathcal{M})
			\end{smallmatrix}
		\end{bmatrix} g_t = \begin{bmatrix}
			\begin{smallmatrix}
				x_t\\
				\bar{u}_{[0,L - 1]|t}\\
				\bar{M}_t
			\end{smallmatrix}
		\end{bmatrix}.
	\end{equation}
	Using the trajectory expression and \eqref{eq:Hg=u:M-thm-pr}, we obtain
	\begin{align*}
		\setlength{\abovedisplayskip}{3pt}
		\setlength{\belowdisplayskip}{3pt} 
		\bar{y}_{[0,L  - 1]|t} & = \mathcal{O}_{L } x_t +  \mathcal{T}^u_{L} \bar{u}_{[0, L - 1]} + \hat{\mathcal{T}}^w_{L} \mathcal{L}(w_0) \mathcal{M}(t)\\
		& \overset{\eqref{eq:Hg=u:M-thm-pr}}{=} \sum_{i = 0}^{N - L} g_{t,i}(\mathcal{O}_{L } x_i + \mathcal{T}^u_{L} u_{[i, i + L  - 1]} + \hat{\mathcal{T}}^w_{L} \mathcal{L}(w_0) \mathcal{M}(i))\\
		& \overset{\eqref{eq:y-sequence-xuw-abbre}}{=} \sum_{i = 0}^{N - L } g_{t,i} y_{[i,i + L - 1]} = H_{L}(y)g_t.
	\end{align*}
	Together with \eqref{eq:Hg=u:M-thm-pr}, this gives
	\eqref{eq:Hg=u:M-no-redundent}, which completes the proof.
\end{proof}

% \begin{remark}
% 	The constraint $H_1(\mathcal{M})g_t=\bar M_t$ fixes the known temporal basis of the exosystem \eqref{eq:sys:w}. The unknown coefficient matrix induced by the measured data is the one appearing in the available trajectory. Hence, the result is applied to predictions along the same exosystem realization, phase convention, or coefficient class represented by the data. If a different exosystem initial condition or amplitude is to be regulated, then the data set must be informative for that realization, or the corresponding exosystem coefficients must be reconstructed and included in the prediction constraint.
% \end{remark}
\begin{remark}[Scope of Theorem~\ref{thm:fund-lem-exo}]
	\label{rem:phase-sync}
	The matrix $\mathcal{L}(w_0)$ in \eqref{eq:H1w=lH1M} is unknown but common to the data and to the predictions, and the constraint $H_1(\mathcal{M})g_t=\bar M_t$ with $\bar M_t=\mathcal M(t)$ selects the exosignal $\bar w_{0|t}=\mathcal L(w_0)\mathcal M(t)=w_t$. Hence, Theorem~\ref{thm:fund-lem-exo} characterizes the trajectories driven by the \emph{same} exosystem realization as the data, evaluated at a known time index $t$. Accordingly, throughout the paper, the exosignal acting during online operation is the continuation $w_t=S^tw_0$ of the realization in the data, and the controller knows the time index $t$ on the time axis of the data, i.e., the phase of the exosystem, but neither $w_0$ nor the samples $w_t$ (the latter are used only in the analytical formulation \eqref{eq:ddmpc-ori} to evaluate the feedforward term). 
	This setting arises in repetitive tasks in which the phase of the exogenous signal is available while its amplitude is not, e.g., path following on a known route, where the road curvature is indexed by the traveled distance; rotating machinery, where periodic disturbances are indexed by the measured shaft angle; and grid-connected converters, where the grid phase is provided by a phase-locked loop. 
	If the online exosignal belongs to a different realization, e.g., with a different initial condition or amplitude, Theorem~\ref{thm:fund-lem-exo} does not apply directly; the data must then be informative for that realization, or the corresponding exosystem coefficients must be reconstructed and included in the prediction constraint.
\end{remark}

Theorem \ref{thm:fund-lem-exo} is the prediction device used in the rest of the paper. It extends the affine-system fundamental lemma in \cite[Theorem 1]{berberich2021linearII} from constant offsets to dynamic exosignals. Indeed, when $S=I$, one has $H_{1}(\mathcal{M})=\bmat{1\cdots1}$ and $\bar{M}_t=1$, so the additional constraint reduces to the usual affine normalization. Thus, \cite[Theorem 1]{berberich2021linearII} is recovered as a special case of Theorem \ref{thm:fund-lem-exo}.

\section{Nominal DDPC for Output Regulation}
\label{sec:ddmpc}

We first consider the ideal case in which the available input-output data and the online measurements are noise-free. The section has two roles. First, it introduces a nominal DDPC formulation whose cost regularizes the input around the steady-state feedforward action, making the connection with model-based output-regulating MPC explicit. Second, for periodic exosignals, it replaces this unavailable feedforward action by an incremental-input penalty, yielding the implementable nominal scheme used later in the robust design.

\subsection{Problem Formulation}\label{sec:ddmpc:prob-form} 

%  \begin{subequations}\label{eq:sys}
	% 	\begin{align}
		%     w_{t + 1} & = S w_t\label{eq:sys:w}\\
		% x_{t + 1} & = Ax_t + Bu_t + Ew_t \label{eq:sys:x}
		% \\
		%  y_t & = C x_t + F w_t\label{eq:sys:e}
		% \end{align}
	%  \end{subequations}
% where $x_t \in \mathbb{R}^{n}$ is the state, $u_t \in \mathbb{R}^{m}$ is the control input, $y_t \in \mathbb{R}^{p}$ is the output, and $w_t \in \mathbb{R}^{n_w}$ represents the exosignal (e.g., external disturbance, and reference input).
% Matrices $A$, $B$, $C$, $E$, and $F$ are unknown constant matrices.
Consider system \eqref{eq:sys} with $m=p$ and $y$ denoting the regulation error. The controller may use measured input-output data, the known exosystem structure, and online measurements, but not the matrices $A$, $B$, $C$, $E$, and $F$. To describe convergence to the regulator manifold, define the extended state $\chi_t:=(x_t,w_t)\in\X\subseteq\R^{n+n_w}$. The corresponding extended dynamics are
\begin{subequations}\label{eq:sys:xi-y}
	\setlength{\abovedisplayskip}{3pt}
	\setlength{\belowdisplayskip}{3pt} 
	\begin{align}
		\chi_{t + 1} & = \underbrace{\smat{A & E\\
				0 & S}}_{:= A_{\chi}}\chi_{t} + \underbrace{\smat{B\\
				0}}_{:= B_{\chi}} u_t\\
		y_t &= \underbrace{\smat{C & F}}_{:= C_{\chi}}\chi_{t}.
	\end{align}
\end{subequations}
The following regulation assumptions are used, consistently with \cite{kohler2021mpc-output-reg}. Assumption \ref{as:reg-eq} is the classical solvability condition for exact regulation; Assumption \ref{as:mat-stab-detect} gives the structural properties needed for finite-memory prediction and stabilization; and Assumption \ref{as:nonresonance-cond} excludes invariant zeros at the exosystem modes.
\begin{assumption}[Output regulator equations]
	\label{as:reg-eq}
	There exist matrices $(\Pi,\Gamma)$ satisfying 
	\begin{subequations}\label{eq:reg-eq}
		\setlength{\abovedisplayskip}{3pt}
		\setlength{\belowdisplayskip}{3pt} 
		\begin{align}
			\Pi S & = A \Pi + B \Gamma + E\\
			0 & = C \Pi  + F.
		\end{align}
	\end{subequations}
	% are solvable.
	% The solution to \eqref{eq:reg-eq} is denoted by $(\Pi,\Gamma)$.
\end{assumption}
\begin{assumption}
	[Controllability and observability]
	\label{as:mat-stab-detect}
	The pair $(A,B)$ is controllable, and the pair $(A,C)$ is observable.
\end{assumption}
\begin{assumption}
	[Nonresonance condition]
	\label{as:nonresonance-cond}
	The following condition holds for all $\lambda \in {\rm spectrum}(S)$
%	{\setlength{\abovedisplayskip}{3pt}
%	\setlength{\belowdisplayskip}{3pt} 
	\begin{equation}\label{eq:nonresonance-cond}
		{\rm rank} \smat{
			A - \lambda I & B\\
			C & 0
		} = n + p.
	\end{equation}
\end{assumption}
\begin{remark}
	The matrix in \eqref{eq:nonresonance-cond} is
	$(n+p)\times(n+m)$, so full row rank requires $m\geq p$. We
	restrict attention to $m=p$; Assumptions \ref{as:reg-eq} and
	\ref{as:nonresonance-cond} then make the regulator pair
	$(\Pi,\Gamma)$ unique. This uniqueness is used later to establish
	detectability of the incremental-input realization. When $m>p$, an
	input-allocation condition is needed to exclude distinct zero-output
	periodic inputs
	\cite[Sec.~V-B]{kohler2021mpc-output-reg}.
\end{remark}
Building on these preliminaries, the data-driven constrained output regulation problem is formally stated as follows.
\begin{problem}\label{prob:orp_ori}
	Consider \eqref{eq:sys} with unknown matrices $A$, $B$, $C$, $E$, $F$ satisfying Assumptions \ref{as:reg-eq}--\ref{as:nonresonance-cond}.
	Given input-output data $\{u_t,y_t\}_{t = 0}^{N - 1}$ generated by \eqref{eq:sys}, and with the online exosignal and the available phase information as specified in Remark~\ref{rem:phase-sync}, design a causal control sequence $u_t$ such that the extended state $\chi_t$ converges to the regulator manifold $\mathcal{A}:=\{\chi\in\X~|~x=\Pi w\}$ and $y_t\to0$ as $t\to\infty$, while satisfying the imposed input constraints.
\end{problem}
Thus the goal is equivalent to driving the distance measure $\sigma(\chi)=\|x-\Pi w\|^2$ to zero. This expression is used for analysis only: in the implementable periodic formulation below, the controller does not require the matrices $\Pi$ and $\Gamma$.

% Although system matrices $A$, $B$, $C$, $E$, $F$ are unknown, some input-output data $u$, $y$ generated from \eqref{eq:sys} can be collected during some offline experiments.
% Hence, in this paper, we aim to solve Problem \ref{prob:orp_ori} using input-output data.

\subsection{Auxiliary State-Space System}
To express regulation in terms of measured input-output histories, we recall the auxiliary realization proposed in \cite[Sec. 2]{bisoffi2026output}. This step replaces the unmeasured plant state by a finite window of past outputs and inputs, i.e., by information available to the data-driven controller.
Since $(A,C)$ is observable (cf. Assumption \ref{as:mat-stab-detect}), let $\iota$ be its observability index. Then
\begin{equation}
	{\rm rank}(\mathcal{O}_\iota) = \smat{C\\
		CA\\
		\vdots\\
		CA^{\iota - 1}} = n.
\end{equation}
Hence there exists matrix $\mathcal{O}^{\dag}$ such that $\mathcal{O}_\iota^{\dag} \mathcal{O}_\iota = I$.
Consider the following dynamics:
\begin{subequations}\label{eq:pre:sys:aux}
	\begin{align}
		\omega_{t + 1} &= \mathbf{S} \omega_t \label{eq:pre:sys:aux:w}\\
		\xi_{t + 1} & = \mathbf{A}_\iota \xi_t + \mathbf{B}_\iota v_t  + \mathbf{L}_\iota Z_2 \omega_t \label{eq:pre:sys:aux:xi}    \\
		\varphi_t & = Z_1 \xi_t + Z_2 \omega_t \label{eq:pre:sys:aux:y}
	\end{align}
\end{subequations}
where $\xi \in \R^{n_\xi}$ with $n_\xi = \iota(m + p)$, $\varphi_t \in \R^{p}$, and 
\begin{align*}
	& \mathbf{S} := {\rm block.diag}(\underbrace{S, \ldots, S}_{(\iota+1) \ \text{times}}),~~\mathbf{A}_\iota := \mathbf{F}_\iota + \mathbf{L}_\iota Z_1,\\
	& \mathbf{F}_\iota := \left[ \begin{array}{c|c}
		\begin{smallmatrix} 
			0 & I_p & 0 & \cdots & 0 \\ 
			0 & 0 & I_p & \cdots & 0 \\ 
			\vdots & \vdots & \vdots & \ddots & \vdots \\ 
			0 & 0 & 0 & \cdots & I_p \\ 
			0 & 0 & 0 & \cdots & 0
		\end{smallmatrix} 
		& 0_{p \iota \times m \iota} \\ 
		\hline 
		0_{m \iota \times p \iota} & 
		\begin{smallmatrix} 
			0 & I_m & 0 & \cdots & 0 \\ 
			0 & 0 & I_m & \cdots & 0 \\ 
			\vdots & \vdots & \vdots & \ddots & \vdots \\ 
			0 & 0 & 0 & \dots & I_m \\ 
			0 & 0 & 0 & \dots & 0
		\end{smallmatrix} 
	\end{array} \right]\\
	& \mathbf{L}_\iota := \left[%
	\begin{array}{c}
		\begin{smallmatrix} 
			0 \\[2pt] 
			0 \\[1pt] 
			\vdots \\[1pt] 
			0 \\[1pt]
			I_p 
		\end{smallmatrix} \\[1pt] 
		\hline  
		\begin{smallmatrix} 
			0 \\[1pt]  
			0 \\[1pt] 
			\vdots \\[1pt] 
			0 \\[1pt] 
			0
		\end{smallmatrix} 
	\end{array} \right], ~
	\mathbf{B}_\iota := \left[ \begin{array}{c}
		\begin{smallmatrix} 
			0 \\[2pt] 
			0 \\[1pt] 
			\vdots \\[1pt] 
			0 \\[1pt]
			0\rotatebox{180}{\rule{0pt}{2.5pt}}
		\end{smallmatrix} \\ 
		\hline  
		\begin{smallmatrix} 
			0 \\[1pt] 
			0 \\[1pt] 
			\vdots \\[1pt] 
			0 \\[1pt] 
			I_m
		\end{smallmatrix}
	\end{array} \right], \\
	&Z_1 := 
	\begin{bmatrix}
		CA^{\iota} \Obs^{\dag} &   C\Ru - CA^{\iota} \Obs^{\dag}\mathcal{T}^u_\iota
	\end{bmatrix},\\
	&Z_2 := 
	\begin{bmatrix}
		C\Rw - CA^{\iota} \Obs^{\tu{L}}\mathcal{T}^w_\iota & Q
	\end{bmatrix}
\end{align*}
where matrices $R_{\rm u}$, $\mathcal{T}^u_\iota$, $R_{\rm w}$, and $\mathcal{T}^w_\iota$ are defined analogously to those in \eqref{eq:matrix_O_T}, see \cite[Section 2]{bisoffi2026output} for details.
It was shown in \cite[Lemma 4.4]{bisoffi2026output} that \eqref{eq:pre:sys:aux} has the same input-output behavior as the original LTI system \eqref{eq:sys}; that is, for $t \ge \iota$, with $v_t = u_t$ and $\omega_t = \smat{w_{t-\iota} \\ \vdots \\ w_t }$, it holds that $	\xi_t = \smat{y_{[t - \iota,t - 1]}\\
	u_{[t - \iota,t - 1]}}$ and $\varphi_t = y_t$.
%\begin{equation}
%	\xi_t = \smat{y_{[t - \iota,t - 1]}\\
%		u_{[t - \iota,t - 1]}},~~ \varphi_t = y_t.
%\end{equation}
In the rest of the paper, we use $u_t := v_t$, $\omega_t := \smat{w_{t-\iota} \\ \vdots \\ w_t }$, $y_t := \varphi_t$, and $\xi_t := \smat{y_{[t - \iota,t - 1]}\\
	u_{[t - \iota,t - 1]}}$ for simplicity. With this notation, \eqref{eq:pre:sys:aux} is rewritten as
\begin{subequations}\label{eq:sys:aux}
	\begin{align}
		\omega_{t + 1} &= \mathbf{S} \omega_t \label{eq:sys:aux:w}\\
		\xi_{t + 1} & = \mathbf{A}_\iota \xi_t + \mathbf{B}_\iota u_t  + \mathbf{E}_\iota \omega_t \label{eq:sys:aux:xi}    \\
		y_t & = Z_1 \xi_t + Z_2 \omega_t \label{eq:sys:aux:y}
	\end{align}
\end{subequations}
where $\mathbf{E}_\iota := \mathbf{L}_\iota Z_2$.
Analogously to \eqref{eq:sys:xi-y}, let $\varrho_t := (\xi_t,\omega_t) \in \X^\varrho  \subseteq \R^{n_\xi + n_w}$ be the extended state of this auxiliary system. Then, \eqref{eq:sys:aux} can be written as
\begin{subequations}\label{eq:sys:aug-varrho}
	\begin{align}
		\varrho_{t + 1} & = \underbrace{\smat{
				\mathbf{A}_\iota & \mathbf{L}_\iota Z_2\\
				0 & \mathbf{S}
		}}_{:= A_{\varrho}}\varrho_{t} + \underbrace{\smat{\mathbf{B}_\iota\\
				0}}_{:= B_{\varrho}} u_t\\
		y_t & = \underbrace{\smat{Z_1 & Z_2}}_{:= C_{\varrho}}\varrho_{t}.
	\end{align}
\end{subequations}
Under Assumption \ref{as:mat-stab-detect}, the homogeneous
(zero-exosignal) auxiliary realization satisfies that
$(\mathbf A_\iota,\mathbf B_\iota)$ is stabilizable and
$(\mathbf A_\iota,Z_1)$ is detectable
\cite[Lemma~1]{berberich2021robust}.
Moreover, Assumption \ref{as:reg-eq} and the input-output
equivalence established in
\cite[Lemma~4.4]{bisoffi2026output} imply that the auxiliary
regulator equations
\begin{subequations}\label{eq:aux-reg-eq}
	\setlength{\abovedisplayskip}{3pt}
	\setlength{\belowdisplayskip}{3pt}
	\begin{align}
		\Pi_\xi\mathbf S
		&=\mathbf A_\iota\Pi_\xi
		+\mathbf B_\iota\Gamma_\xi+\mathbf E_\iota,\\
		0&=Z_1\Pi_\xi+Z_2
	\end{align}
\end{subequations}
admit a solution, denoted by $(\Pi_\xi,\Gamma_\xi)$.
% To avoid notational burden,
With
\begin{equation}\label{eq:sigma-varrho}
	\sigma(\varrho) := \Vert \xi - \Pi_\xi \omega\Vert^2 = \Vert e_\xi\Vert^2
\end{equation}
the output regulation problem for the auxiliary system \eqref{eq:sys:aux} is stated as follows.
\begin{problem}\label{prob:orp_aux}
	Consider system \eqref{eq:sys:aux} with unknown matrices $\mathbf{A}_\iota$, $\mathbf{B}_\iota$, $\mathbf{E}_\iota$, $Z_1$, and $Z_2$, where the underlying plant \eqref{eq:sys} satisfies Assumptions \ref{as:reg-eq}--\ref{as:nonresonance-cond}, so that $(\mathbf A_\iota,\mathbf B_\iota)$ is stabilizable and $(\mathbf A_\iota,Z_1)$ is detectable.
	Given input-output data $\{u_t,y_t\}_{t = 0}^{N - 1}$ generated by \eqref{eq:sys}, find a control sequence $u$ that drives $\varrho_t$ to the regulator manifold \( \mathcal{A}_\varrho := \{ \varrho \in \mathcal{X}^\varrho \mid \xi = \Pi_\xi \omega \} \) and satisfies $y_t \rightarrow 0$ as $t \rightarrow \infty$.
\end{problem}
% If we can design a sequence of control input \( u\) such that \( \sigma(\varrho) = 0 \), then the overall state \( \varrho \) is driven to the regulator manifold \( \mathcal{A}_\varrho := \{ \varrho \in \mathcal{X}^\varrho \mid \xi = \Pi_\xi \omega \} \). 
Since systems \eqref{eq:sys} and \eqref{eq:sys:aux} have the same input-output behavior, any input solving Problem \ref{prob:orp_aux} also solves Problem~\ref{prob:orp_ori}.
% \begin{subequations}
	%     \begin{align}
		%         &\bar{c}_{o,l} \Vert \xi - \bar{\xi}\Vert^2 \le \bar{V}_o(\xi,\bar{\xi},\omega) \le \bar{c}_{o,u} \Vert \xi - \bar{\xi}\Vert^2 \label{eq:low-up-bound-Vo}\\
		%         &\bar{V}_o(\xi_{t + 1}, \bar{\xi}_{t + 1},\textbf{S}\omega) - \bar{\rho}_o V_o(\xi_{t}, \bar{\xi}_{t},\omega) \le \bar{c}_{o,1}\Vert \bar{v} - v\Vert^2 + \bar{c}_{o,2}\Vert Z_1\xi_t - Z_1\bar{\xi}_t )\Vert^2 \label{eq:decrease-Vo}
		%     \end{align}
	% \end{subequations} 
% where $\bar{V}_o(\xi,\bar{\xi},\omega) := \Vert \xi - \bar{\xi} \Vert_{\bar{P}_o}^2$ is an i-IOSS Lyapunov function.
% Let $W(\xi_t) = \Vert \xi_t - \Pi \omega_t\Vert^2_{P_V}$ with $c = \min\left\{\frac{\lambda_{\min}(R)}{c_{o,1}}, \frac{\lambda_{\max}(Q)}{c_{o,2}}\right\} > 0$.
We now present the nominal DDPC formulation derived from the trajectory parametrization above.
% In the following, we study the closed-loop properties of a simple nominal DDPC for output regulation.
% First, a nominal DDPC scheme solving Problem \ref{prob:orp_ori} is developed in the case of noise-free measurement and data, with knowledge of the input feedforward $\Gamma_\xi \omega$.

% Building on this framework, we consider periodic exosignal $w$, and replace the feedforward regularization term with an incremental input regularization term, followed by exponential stability guarantees.
% In this case, 
% Next, for periodic exosignal, 
% This scheme may be rather restrictive and only serves as a preliminary to the incremental input regularization and robustness results in Sections \ref{sec:ddmpc-incre} and \ref{sec:ddmpc-noise}.

\subsection{Nominal DDPC}
\label{sec:ddmpc:mpc-scheme}
For clean input-output data, the prediction model of Theorem \ref{thm:fund-lem-exo} can be incorporated into a receding-horizon regulator. The resulting optimization problem is
\begin{subequations}\label{eq:ddmpc-ori}
	\begin{align}
		\min_{\bar{y},\bar{u},\bar{\omega},g}~~& J_L(\xi_t, \omega_t)= \sum_{k = 0}^{L- 1} \Vert \bar{y}_{k|t}\Vert^2_{Q} + \Vert \bar{u}_{k|t} - \Gamma_\xi \bar{\omega}_{k|t}\Vert^2_{R}  \label{eq:ddmpc-ori:obj} \\
		{\rm s.t.}~~&\begin{bmatrix}
			\begin{smallmatrix}
				H_{L + \iota}(y)\\
				H_{L + \iota}(u)\\
				H_1(\mathcal{M})
			\end{smallmatrix}
		\end{bmatrix} g_t = \begin{bmatrix}
			\begin{smallmatrix}
				\bar{y}_{[-\iota,L - 1]|t}\\
				\bar{u}_{[-\iota,L - 1]|t}\\
				\bar{M}_{t - \iota}
			\end{smallmatrix}
		\end{bmatrix} \label{eq:ddmpc-ori:hankel}\\
		& \begin{bmatrix}
			\begin{smallmatrix}
				\bar{y}_{[-\iota,-1]|t}\\
				\bar{u}_{[-\iota,-1]|t}
			\end{smallmatrix}
		\end{bmatrix} = \begin{bmatrix}
			\begin{smallmatrix}
				y_{[t - \iota, t - 1]}\\
				u_{[t - \iota,t - 1]}
			\end{smallmatrix}
		\end{bmatrix} \label{eq:ddmpc-ori:init-cond}\\
		&\bar{\omega}_{0|t} = \smat{{w}_{t - \iota}\\
			\vdots\\
			{w}_{t}} \label{eq:ddmpc-ori:init-cond-w}\\
		&\bar{\omega}_{k + 1|t} = \mathbf{S} \bar{\omega}_{k|t}, ~~~k = 0, \cdots, L - 1 \label{eq:ddmpc-ori:w}\\
		& \bar{u}_{k|t} \in \mathbb{U}, ~~~k = 0, \cdots, L - 1\label{eq:ddmpc-ori:u-bound-cond}
	\end{align}
\end{subequations}
where $L \in \mathbb{N}$ is the prediction horizon and $\mathbb{U}$ is a compact input-constraint set. The decision variables are the predicted output, input, exosystem basis trajectory, and Hankel coefficient vector. The constraints and cost have the following interpretation.
\begin{itemize}
	\item The stage cost in \eqref{eq:ddmpc-ori:obj}, i.e.,
	\begin{equation}
		\setlength{\abovedisplayskip}{3pt}
		\setlength{\belowdisplayskip}{3pt} 
		\label{eq:stagy_cost_ell}
		\ell(\xi,\omega) := \Vert y\Vert_Q^2 + \Vert u - \Gamma_\xi \omega \Vert^2_R 
	\end{equation} 
	penalizes both the regulation error and the deviation of the predicted input from the steady-state feedforward action $\Gamma_\xi\omega$, with $Q=Q^\top\succ0$ and $R=R^\top\succ0$.
	\item 
	Constraint \eqref{eq:ddmpc-ori:hankel} is a data-driven prediction model supplied by Theorem \ref{thm:fund-lem-exo}; it replaces the state-space equations used in the model-based MPC of \cite[eq. (5)]{kohler2021mpc-output-reg}.
	\item Constraints \eqref{eq:ddmpc-ori:init-cond}--\eqref{eq:ddmpc-ori:init-cond-w} ensure that the internal state of the true trajectory aligns with the internal state of the predicted trajectory at time $t$. 
\end{itemize}
The cost \eqref{eq:stagy_cost_ell} also reveals the main limitation of \eqref{eq:ddmpc-ori}: implementation requires the feedforward signal $\Gamma_\xi\omega$.
% To solve the optimization problem \eqref{eq:ddmpc-ori} with , we need to be able to predict both the regulation error $y$ and the state of the exosystem $\omega$, and thus we initially assume that the overall state $(\xi,\omega)$ can be measured.
This requirement is restrictive because $\Gamma_\xi$ is tied to the unknown regulator equations. Therefore, \eqref{eq:ddmpc-ori} serves as an analytical bridge to existing output-regulating MPC results; the feedforward requirement is removed in Section \ref{sec:ddmpc:incre-mpc}.

%\begin{remark}[Why not standard DDPC]
%	Applying the standard fundamental lemma directly to the augmented state
%	$\chi=(x,w)$ is not possible because the exosystem evolves autonomously
%	and $(A_\chi,B_\chi)$ is generally uncontrollable. Moreover, a Hankel
%	constraint involving only $u$ and $y$ does not enforce consistency with
%	the required exosystem basis vector. Theorem~\ref{thm:fund-lem-exo}
%	addresses this issue by imposing
%	$H_1(\mathcal M)g_t=\bar M_t$.
%	
%	Even with the predictor of Theorem~\ref{thm:fund-lem-exo}, the usual
%	equilibrium-centered DDPC stage cost
%	$\|y\|_Q^2+\|u\|_R^2$ is not positive definite with respect to the
%	regulator manifold. On this manifold, $y=0$, whereas regulation
%	generally requires the nonzero, time-varying input
%	$u=\Gamma_\xi\omega$. Penalizing $u$ would therefore bias the optimizer
%	away from the regulated motion.
%\end{remark}

% with a prediction horizon $L \in \mathbb{N}$,
% For simplicity, we assume $\mathbb{U}$ compact.
When Problem \eqref{eq:ddmpc-ori} is feasible at time $t$, denote its optimizer by $J_L^*$, $(\bar{y}^*_t,\bar{u}^*_t,\bar{\omega}^*_t)$, and $g_t^*$.
The resulting receding-horizon implementation is summarized in Algorithm \ref{alg:ddmpc-ori}.
\begin{algorithm}[h]
	\small
	\caption{Nominal DDPC for Output Regulation.}
	\label{alg:ddmpc-ori}
	\begin{algorithmic}[1]
		\State {\bfseries Given:} Prediction horizon $L \ge  \iota + {n}$, I/O data $\{u_t,y_t\}_{t = 0}^{N - 1}$ that are persistently exciting of order $L + \iota$ in the sense of Definition~\ref{def:pe}, the constraint set $\mathbb{U}$, feedforward matrix $\Gamma_\xi$, and I/O weighting matrices $Q$, $R$.
		
		\State {\bfseries If} $t \le \iota - 1$, do \label{alg:ddmpc-ori-step1}				
		\State \quad Run \eqref{eq:sys} in open loop with randomly chosen input $u_t$, and store measurements $u_t$, $y_t$, $w_t$.
		
		\State {\bfseries Else if} $t \ge \iota$, do 
		% \STATE  \quad 
		%          Take the past exosignal $w_{t - \iota}$ and solve \eqref{eq:hg=w-ss} to obtain $g^s_t$.
		\State  \quad Take the past $\iota$ I/O measurements $(u_{[t - \iota, t - 1]}, y_{[t - \iota, t - 1]})$ and exosignal samples $w_{[t - \iota,t]}$ (used only in \eqref{eq:ddmpc-ori:init-cond-w} to evaluate the feedforward term), set $\bar{M}_{t - \iota}=\mathcal{M}(t-\iota)$, with $t$ counted on the time axis of the data (cf. Remark~\ref{rem:phase-sync}), and solve \eqref{eq:ddmpc-ori}.
		
		\State \quad Apply the input $u_t = \bar{u}^*_{0|t}$.
		
		\State \quad  Set $t := t + 1$ and go back to \ref{alg:ddmpc-ori-step1}.
	\end{algorithmic}
\end{algorithm}

The following result gives the closed-loop guarantee associated with Algorithm \ref{alg:ddmpc-ori}.
% The resulting closed-loop system is given by 
% \begin{equation}
	%     u_t = \bar{u}^*_{0|t},~~w_{t + 1} = S w_t,~~x_{t + 1} = A x_t + B \bar{u}^*_{0|t} + E w_t,~~y_t = C x_t + F w_t = \bar{y}^*_{0|t},~~t \ge 0.
	% \end{equation}

\subsection{Closed-Loop Stability Guarantees}
\label{sec:ddmpc:stab-analy}
Before proceeding, we impose a uniform interior condition on the
regulator input. It provides the input margin needed for the local
feasibility argument below.
\begin{assumption}[Admissibility of regulator input]
	\label{as:strict-reg-input}
	For the exosignal $\omega_t$ in \eqref{eq:sys:aux:w}, letting $	\mathbb B_{r_{\mathbb U}}(\Gamma_\xi\omega_t)
	:=\bigl\{v\in\mathbb R^m\mid
	\|v-\Gamma_\xi\omega_t\|\le r_{\mathbb U}\bigr\}$, there exists
	$r_{\mathbb U}>0$ such that $\mathbb B_{r_{\mathbb U}}(\Gamma_\xi\omega_t)
	\subseteq\mathbb U$ for all $t\ge0$.
%	\begin{equation}\label{eq:strict-reg-input}
%		\mathbb B_{r_{\mathbb U}}(\Gamma_\xi\omega_t)
%		\subseteq\mathbb U,~~ \forall t\ge0.
%	\end{equation}
\end{assumption}
\color{black}
% with the i-IOSS Lyapunov function $W(\xi_t,\omega_t) := \Vert \xi_t - \Pi_\xi \omega_t \Vert_{{P}_{\xi,o}}^2$.
We next recall two ingredients from \cite[Section V.D]{kohler2021mpc-output-reg}. Under Assumptions \ref{as:reg-eq}--\ref{as:strict-reg-input}, the following bounds hold for the corresponding output-regulating MPC problem.
\begin{proposition}
	\label{prop:stab}
	Under Assumptions \ref{as:mat-stab-detect} and \ref{as:strict-reg-input}, there exist constants $\gamma_s, \delta_s >0$, such that for any $\varrho \in \mathbb{X}^{\varrho}_{\delta} := \{\varrho \in \X^{\varrho} | \sigma(\varrho) \le \delta_s \}$ with $\sigma(\varrho)$ defined in \eqref{eq:sigma-varrho}, problem \eqref{eq:ddmpc-ori} is feasible and the objective function satisfies 
	% \begin{equation}\label{eq:up-bound-V}
		%     V_N(\chi) \le \gamma_s \sigma(\chi),~\forall N \in \mathbb{N}.
		% \end{equation}
%		{\setlength{\abovedisplayskip}{3pt}
%		\setlength{\belowdisplayskip}{3pt}
	\begin{equation}\label{eq:up-bound-J}
		J^*_L(\xi_t,\omega_t) \le \gamma_s \Vert \xi_t - \Pi_\xi \omega_t\Vert^2. 
	\end{equation}
\end{proposition}
% \begin{proposition}\label{prop:detect}
	%     Under Assumption \ref{as:mat-stab-detect}, there exists function $W \in \X \rightarrow \R_{\ge 0}$ and constants $\gamma_o, \epsilon_o >0$, such that 
	%     \begin{subequations}\label{eq:up-bound-decrease-W}
		%         \begin{align}
			%             W(\chi) & \le \gamma_o \sigma(\chi) \label{eq:up-bound-W}\\
			%             W(A_\chi \chi + B_\chi u) - W(\chi) & \le - \epsilon_o \sigma(\chi_t) + \ell(\chi,u)\label{eq:iioss-decrease-W}
			%         \end{align}
		%     \end{subequations}
	% \end{proposition}
% \begin{proposition}\label{prop:i-IOSS}
	%     Under Assumption \ref{as:mat-stab-detect}, there exist a continuous incremental input-output-to-state (i-IOSS) Lyapunov function $V_o : \R^{n} \times \R^{n} \times \W \rightarrow R_{\ge 0}$ and constants $c_{o,l}, c_{o,u}, c_{o,1}, c_{o,2} > 0$, $\rho_o \in (0,1)$, such that for any $(z,v,w,x,u)$, we have 
	%     \begin{subequations}
		%         \begin{align}
			%             &c_{o,l} \Vert x - z\Vert^2 \le V_o(x,z,w) \le c_{o,u} \Vert x - z\Vert^2 \label{eq:low-up-bound-Vo}\\
			%             &V_o(Ax + Bu + Ew, Az + Bv + Ew,Sw) - \rho_o V_o(x,z,w) \le c_{o,1}\Vert u - v\Vert^2 + c_{o,2}\Vert Cx + Fw - (Cz + F w)\Vert^2. \label{eq:decrease-Vo}
			%         \end{align}
		%     \end{subequations}
	% \end{proposition}
\begin{proposition}\label{prop:i-IOSS}
	Under Assumptions \ref{as:reg-eq} and \ref{as:mat-stab-detect}, system \eqref{eq:sys:aug-varrho} is incrementally input-output-to-state stable (i-IOSS) \cite[Definition 4]{allan2019lyapunov}. In addition, there exist matrix ${P}_{\xi,W}=P_{\xi,W}^\top \succ 0$ and constant $\epsilon_o > 0$ such that, with $W(\varrho_t)=\Vert \xi_t-\Pi_\xi\omega_t\Vert^2_{P_{\xi,W}}$,
	the following inequalities hold for all $(\varrho_t,u_t)$:
	\begin{subequations}\label{eq:iioss-bounds-W}
		\begin{align}
			W(\varrho_t) &\ge \epsilon_o \Vert \xi_t-\Pi_\xi\omega_t\Vert^2, \label{eq:iioss-lower-W}\\
			W(\varrho_{t + 1}) &\le W(\varrho_t) + \Vert u_t - \Gamma_\xi  \omega_t\Vert^2_R + \Vert y_t \Vert_{Q}^2\nonumber\\
			&\quad - \epsilon_o \Vert \xi_t - \Pi_\xi \omega_t \Vert^2.
			\label{eq:iioss-decrease-W}
		\end{align}
	\end{subequations}
\end{proposition}

For the stability analysis of Algorithm \ref{alg:ddmpc-ori}, consider the Lyapunov function $ Y_{L}(\varrho_t) := J^*(\varrho_t) + W(\varrho_t)$.
% Consider the following continuous state measure to be minimized 
% \begin{equation}\label{eq:sigma-chi}
	%     \sigma(\varrho) := \Vert \xi - \Pi_\xi \omega\Vert^2 = \Vert e_\xi\Vert^2.
	% \end{equation}
% Achieving $\sigma(\varrho) = 0$ equivalent to driving the overall state to the regulator manifold $\mathcal{A}_\varrho := \{\xi \in \X^{\varrho}|\xi = \Pi_\xi \omega \}$.
% This further implies $\sigma(\chi) := \Vert x - \Pi w \Vert^2 = \Vert e\Vert^2 = 0$, and hence solves the output regulation problem of the system \eqref{eq:sys}.
%The next theorem transfers the model-based result in \cite[Theorem 1]{kohler2021mpc-output-reg} to the data-driven formulation through Theorem \ref{thm:fund-lem-exo}.
The following theorem establishes the closed-loop stability of the proposed
data-driven formulation. The proof combines the trajectory equivalence in
Theorem~\ref{thm:fund-lem-exo} with the finite-horizon Lyapunov framework
for output-regulating MPC in
\cite[Theorem~1]{kohler2021mpc-output-reg}.

\begin{theorem}
	[Nominal stability guarantees]
	\label{thm:stab-ori-ddmpc}
	Suppose Assumptions \ref{as:reg-eq}--\ref{as:strict-reg-input} hold and the I/O data are persistently exciting of order $L+\iota$ in the sense of Definition~\ref{def:pe}.
	Then, for any $\bar{Y}>0$, there exists prediction horizon $L_{\bar{Y}}>0$ such that, for all $L>L_{\bar{Y}}$ and all initial conditions $\varrho_0 \in \X_{\bar{Y}}^{\varrho}:=\{\varrho_0\in\X^\varrho~|~Y_L(\varrho_0)\le\bar{Y}\}$, the DDPC problem \eqref{eq:ddmpc-ori} is recursively feasible and the regulator manifold $\mathcal{A}_\varrho$ is exponentially stable for the closed loop.
\end{theorem}
\begin{proof}
	See Appendix \ref{app:proof-thm2}.
%	See Appendix A in the Supplementary File.
\end{proof}

%The intermediate bound \eqref{eq:iteration-YW} and the definition of $\alpha_L$ in \eqref{eq:alpha_N} imply that as $L \rightarrow \infty$, we recover $\sum_{t = 0}^\infty \ell(\varrho_t) \le J_{\infty}(\varrho_t)$, similar to standard results with $\ell$ positive definite.
%
%
%
%
%
%
%Overall, we have shown that the nominal MPC scheme solves the linear output regulation problem if 
%\begin{itemize}
%	\item The regulator problem is feasible (cf. Assumption \ref{as:reg-eq}),
%	\item 
%	The system \eqref{eq:sys} is controllable and observable (cf. Assumption \ref{as:mat-stab-detect}), and 
%	\item The regulator input is uniformly strictly admissible (cf. Assumption \ref{as:strict-reg-input}).
%\end{itemize}

The controller \eqref{eq:ddmpc-ori} requires the feedforward input $\Gamma_\xi\omega$, and hence either knowledge of the regulator gain or direct access to the regulated steady-state input. This is undesirable in a data-driven setting. We now focus on periodic exosignals and replace feedforward regularization by incremental-input regularization, which is computable from past applied inputs.

\subsection{Incremental Input Regularization for Periodic Exosignals}
\label{sec:ddmpc:incre-mpc}

The periodic setting is captured by the following assumption.

% \begin{assumption}[Periodic exosignal]
	% \label{as:period_exosignal}
	%     There exists known period length $T \in \mathbb{N}$, such that $w_{t + T} = w_t$ for all $t \ge 0$.
	%     In addition, the exosignal $w$ evolves according to $\eqref{eq:sys:w}$, where matrix $S$ is known.
	% \end{assumption}

\begin{assumption}[Periodic exosystem]
	\label{as:period_exosignal}
	The exosystem matrix $S$ in \eqref{eq:sys:w} is known and has finite order
	$T\in\mathbb{N}_{+}$, i.e., $S^T=I$.
\end{assumption}
\begin{remark}
	Assumption~\ref{as:period_exosignal} implies that every trajectory generated
	by $w_{t+1}=Sw_t$ is $T$-periodic. Equivalently, $S$ is diagonalizable over
	$\mathbb{C}$ and all its eigenvalues are $T$-th roots of unity. Thus, for a
	real-valued exosystem, the eigenvalues of $S$ consist of $1$, possibly $-1$,
	and complex-conjugate pairs $e^{\pm \mathrm{i}2\pi q/T}$, with no nontrivial
	Jordan blocks.
	
	The same periodic formulation also covers the case where the particular
	linear matrix $S$ is not known, or where the exosignal is generated by a
	nonlinear periodic map, provided that its period $T$ is known. Indeed, for any
	$T$-periodic sequence $\{w_t\}$, define the lifted exosystem state
	\[
	\mathbf w_t :=
	\smat{
		w_{t-T+1}\\
		w_{t-T+2}\\
		\vdots\\
		w_t},
	\]
	and the cyclic-shift matrix
	\[
	S_T :=
	\smat{
		0 & I_{n_w} & 0 & \cdots & 0\\
		0 & 0 & I_{n_w} & \cdots & 0\\
		\vdots & \vdots & \vdots & \ddots & \vdots\\
		0 & 0 & 0 & \cdots & I_{n_w}\\
		I_{n_w} & 0 & 0 & \cdots & 0
	},
	~~
	S_{\rm sel}:=
	\smat{
		0 & 0 & \cdots & I_{n_w}
	}.
	\]
	Then $\mathbf w_{t+1}=S_T\mathbf w_t$, $S_T^T=I$, and
	$w_t=S_{\rm sel}\mathbf w_t$. Hence \eqref{eq:sys} can be equivalently written as
	\begin{align*}
		\setlength{\abovedisplayskip}{3pt}
		\setlength{\belowdisplayskip}{3pt} 
		x_{t+1} &= A x_t + B u_t + E S_{\rm sel}\mathbf w_t,\\
		y_t &= C x_t + F S_{\rm sel}\mathbf w_t,\\
		\mathbf w_{t+1} &= S_T\mathbf w_t.
	\end{align*}
	Therefore, for periodic exosignals, knowledge of the period is sufficient to
	construct a known linear periodic exosystem.
\end{remark}
\color{black}

\begin{assumption}[Nonresonance at the $T$-th roots of unity]
	\label{as:nonresonance-T}
	Condition \eqref{eq:nonresonance-cond} holds for all $\lambda\in\mathbb{C}$ satisfying $\lambda^T=1$.
\end{assumption}
\begin{remark}
	Under Assumption~\ref{as:period_exosignal}, every eigenvalue of $S$ satisfies $\lambda^T=1$, so Assumption~\ref{as:nonresonance-T} implies Assumption~\ref{as:nonresonance-cond}. The strengthening is needed only for the incremental-input formulation below. The input memory used to form $u_t-u_{t-T}$ has all $T$-th roots of unity as eigenvalues, including those that are not eigenvalues of $S$. An input component oscillating at such a frequency is not penalized by the incremental-input cost, and it must therefore be detectable through the regulated output, cf. the proof of Proposition~\ref{prop:aug-ctrl-detect-equ-ori-ctrl-detect}. Assumption~\ref{as:nonresonance-T} fails only if the plant has an invariant zero located exactly at a $T$-th root of unity; hence, it is generic, and it holds in the example of Section~\ref{sec:numerical_example}.
\end{remark}

% \begin{remark}[Periodic exosignal]
	%     If the period $T$ is known, the same construction can be applied even when the particular matrix $S$ is unknown, or when the exosignal is generated by a nonlinear periodic map $w_{t+1}=\sigma_s(w_t)$. Indeed, define
	% \begin{align*}
		%    &\mathbf{w}_t = \begin{bmatrix}
			%        \begin{smallmatrix}
				%            w_{t - T + 1}\\
				%            w_{t - T + 2}\\
				%            \vdots\\
				%            w_t
				%        \end{smallmatrix}
			%    \end{bmatrix}, ~~
		%    S_T = \begin{bmatrix}
			%         \begin{smallmatrix}
				%             0 & I_{n_w} & 0 & \ldots & 0\\
				%             0 & 0 & I_{n_w} & \ldots & 0\\
				%             \vdots & \vdots & \vdots & \ddots & \vdots\\
				%            I_{n_w}  & 0 & 0 & \ldots &  0
				%         \end{smallmatrix}
			%     \end{bmatrix} \in \mathbb{R}^{n_w T \times n_w T},\\ 
		%     &S_{\mathrm{sel}} = \begin{bmatrix}
			%         \begin{smallmatrix}
				%             0 & 0 & \ldots & I_{n_w}
				%         \end{smallmatrix}
			%     \end{bmatrix} \in \mathbb{R}^{n_w \times n_w T}
		% \end{align*}
	% Then Assumption~\ref{as:period_exosignal} implies $\mathbf{w}_{t+1}=S_T\mathbf{w}_t$, and \eqref{eq:sys} can be rewritten as
	% \begin{subequations}
		%     \begin{align}
			%         x_{t + 1} &= A x_t + B u_t + E S_{\mathrm{sel}} \mathbf{w}_t, \\
			%         y_t &= C x_t + F S_{\mathrm{sel}} \mathbf{w}_t, \\
			%         \mathbf{w}_{t + 1} &= S_T\mathbf{w}_t.
			%     \end{align}
		% \end{subequations}
	% \end{remark}

% Define a memory state for the past applied inputs as $\zeta_t = [u_{t-1}^\top~\cdots~u_{t - T}^\top]^\top \in \R^{mT}$.
Under Assumption \ref{as:period_exosignal}, the steady-state feedforward input is periodic. At regulation one therefore has $u_t=u_{t-T}$, so the unavailable term $\Gamma_\xi\omega$ can be replaced by the measurable periodic input increment $u_t-u_{t-T}$. The resulting data-driven counterpart of \cite[eq. (18)]{kohler2021mpc-output-reg} is
\begin{subequations}\label{eq:ddmpc-incre}
	\setlength{\abovedisplayskip}{3pt}
	\setlength{\belowdisplayskip}{3pt}
	\begin{align}
		\min_{\bar{y},\bar{u},g}~~& J^a_L(\xi_t, \zeta_t)= \sum_{k = 0}^{L- 1} \Vert \bar{y}_{k|t}\Vert^2_{Q} + \Vert \Delta\bar{u}_{k|t}\Vert^2_{R}  \label{eq:ddmpc-incre:obj} \\
		{\rm s.t.}~~&\begin{bmatrix}
			\begin{smallmatrix}
				H_{L + \iota}(y)\\
				H_{L + \iota}(u)\\
				H_1(\mathcal{M})
			\end{smallmatrix}
		\end{bmatrix} g_t = \begin{bmatrix}
			\begin{smallmatrix}
				\bar{y}_{[-\iota, L  - 1]|t}\\
				\bar{u}_{[-\iota, L - 1]|t}\\
				\bar{M}_{t - \iota}
			\end{smallmatrix}
		\end{bmatrix}  \label{eq:ddmpc-incre:hankel}\\
		& \begin{bmatrix}
			\begin{smallmatrix}
				\bar{y}_{[-\iota,-1]|t}\\
				\bar{u}_{[-\iota,-1]|t}
			\end{smallmatrix}
		\end{bmatrix} = \begin{bmatrix}
			\begin{smallmatrix}
				y_{[t - \iota, t - 1]}\\
				u_{[t - \iota,t - 1]}
			\end{smallmatrix}
		\end{bmatrix}  \label{eq:ddmpc-incre:init-cond}\\
		% & \begin{bmatrix}
			%     \begin{smallmatrix}
				%         \bar{y}_{L - n, L - 1}(t)\\
				%         \bar{u}_{L - n, L - 1}(t)
				%     \end{smallmatrix}
			% \end{bmatrix} = \begin{bmatrix}
			%     \begin{smallmatrix}
				%         0\\
				%         0
				%     \end{smallmatrix}
			% \end{bmatrix}  \label{eq:ddmpc-incre:delta-u}\\
		& \bar{u}_{k|t} = \bar{u}_{k - T|t} + \Delta\bar{u}_{k|t},~~~k=0,\cdots,L-1  \label{eq:ddmpc-incre:delta-u}\\
		& \bar{u}_{k - T|t} = \begin{cases}
			u_{t + k - T},&~~k < T\\
			\bar{u}_{k - T|t}, &~~k \ge T
		\end{cases}
		\label{eq:ddmpc-incre:period-cond-u}\\
		& \bar{u}_{k|t} \in \mathbb{U},~~~k=0,\cdots,L-1\label{eq:ddmpc-incre:bound-cond-u}.
	\end{align}  
\end{subequations}

The key point here is that \eqref{eq:ddmpc-incre} is not merely a heuristic replacement of the steady-state feedforward term. As in \cite[Section IV.B]{kohler2021mpc-output-reg}, it is equivalent to a feedforward-regularized MPC problem posed on an augmented plant whose input is the periodic increment.

\emph{Augmented plant}: Define an input memory state $\zeta_t := [u_{t-1}^\top~\cdots~u_{t - T}^\top]^\top \in \R^{mT}$ and set $n_a := n_{\xi} + mT$. Consider
\begin{equation}\label{eq:sys:aug-xu}
	\xi_t^{a} = (\xi_t,\zeta_t) \in \R^{n_a},~~u_t^a = u_t - u_{t - T} \in \R^m.
\end{equation}
Define the block cyclic permutation matrix $E_0 \in \mathbb{R}^{mT \times mT}$ and the selection matrices $E_1,E_2 \in \mathbb{R}^{mT \times m}$:
\begin{equation*}
	E_0 := \smat{0_m & \cdots & 0_m & I_m\\
		I_m & \cdots & 0_m & 0_m\\
		\vdots & \ddots & \vdots & \vdots\\
		0_m & \cdots & 0_m & 0_m\\
		0_m & \cdots & I_m & 0_m  
	},~E_1 := \smat{I_m\\
		0_m\\
		\vdots\\
		0_m},~E_2 := \smat{0_m\\
		\vdots\\
		0_m\\
		I_m}.
\end{equation*}
This matrix satisfies $E_0^\top E_0 = I_{mT}$ and the eigenvalues of $E_0$ are $\lambda_k = e^{{\rm i}2\pi k/T}$, $k = 0, \cdots, T - 1$, all with a geometric and algebraic multiplicity of $m$ (due to the block structure).
We note that $E_1^\top E_0 = E_2^\top$.
The dynamics of the memory state $\zeta$ and the input $u$ can be compactly expressed as 
\begin{equation}\label{eq:sys:aug-xi}
	\setlength{\abovedisplayskip}{3pt}
	\setlength{\belowdisplayskip}{3pt} 
	\zeta_{t + 1} = E_0 \zeta_t + E_1 u_t^a,~~u_t = E_2^\top \zeta_t + u_t^a.
\end{equation}
Thus, the augmented plant dynamics becomes 
\begin{subequations}\label{eq:sys:aug}
	\begin{align}
		\omega_{t + 1} & = \mathbf{S}\omega_t \label{eq:sys:aug-omega}\\
		\xi_{t + 1}^a & = \smat{\mathbf{A}_\iota & \mathbf{B}_\iota E_2^\top\\
			0 & E_0}\smat{\xi_t\\
			\zeta_t} + \smat{\mathbf{B}_\iota\\
			E_1} u_t^a + \smat{\mathbf{E}_\iota\\
			0} \omega_t \nonumber\\
		& := A^a \xi^a_t + B^a u_t^a + E^a \omega_t \label{eq:sys:aug-xa}\\
		y_t &= \smat{Z_1 & 0}\xi^a_t + Z_2 \omega_t := C^a \xi_t^a + Z_2 \omega_t \label{eq:sys:aug-y}.
	\end{align}
\end{subequations}
The overall state of this augmented plant is $\varrho^a := (\xi^a,\omega)$.
%The following lemma, adapted from \cite[Lemma 1]{kohler2021mpc-output-reg}, transfers the feedforward-regularized formulation to the incremental formulation \eqref{eq:ddmpc-incre}.
Using the periodic input-memory augmentation motivated by
\cite[Lemma~1]{kohler2021mpc-output-reg}, the feedforward-regularized
DDPC can be equivalently expressed in the implementable incremental form \eqref{eq:ddmpc-incre}.
\begin{lemma}\label{lem:aug-mpc-equ-ori-mpc}
	The DDPC problem \eqref{eq:ddmpc-ori} with the augmented state $\varrho^a = (\xi^a,\omega)$, input $u^a$, the dynamics \eqref{eq:sys:aug-xu}-\eqref{eq:sys:aug-y}, and stage cost $\ell^a(\varrho^a,u^a) = \Vert y\Vert^2_Q + \Vert u^a\Vert_R^2$ using incremental input regularization is equivalent to the DDPC problem \eqref{eq:ddmpc-incre}.
\end{lemma}

%It remains to verify that the augmented system satisfies the same structural assumptions. 
It remains to verify that the augmented realization retains the structural
properties required for the stability analysis. These properties are
established next using the PBH conditions for the periodic input-memory
augmentation (cf. \cite[Propositions~5 and~7]{kohler2021mpc-output-reg}).
%The proof follows the argument of \cite[Propositions 5 and 7]{kohler2021mpc-output-reg}.
\begin{proposition}\label{prop:aug-ctrl-detect-equ-ori-ctrl-detect}
	Let Assumptions \ref{as:reg-eq}--\ref{as:nonresonance-T} hold.
	The triplet $(A^a, B^a, C^a)$ of the augmented system \eqref{eq:sys:aug-xu}-\eqref{eq:sys:aug} is stabilizable and detectable, and the augmented regulator equations $\Pi^a\mathbf S
	=
	A^a\Pi^a+B^a\Gamma^a+E^a$, $0=C^a\Pi^a+Z_2$
	are solvable.
\end{proposition}
\begin{proof}
	See  Appendix \ref{app:proof-prop4}.
%See Appendix B in the Supplementary File.
\end{proof}
% According to \cite[Propertition 8]{kohler2021mpc-output-reg}, Assumption \ref{as:nonresonance-cond} is equivalent to the following assumption.
% \begin{assumption}
	%     There exist a continuous incremental storage function $V_R: \R^{n_a} \times \R^{n_a} \times \R^{n_w} \mapsto\R_{\ge 0}$ and constants $c_{R,u},c_R >0$, such that for any $(z^a,v^a,x^a,u^a,w)$, the following relationships hold
	%     \begin{subequations}
		%         \begin{align}
			%             & V_R(x^a,z^a,w) \le c_{R,u}\Vert x^a - z^a\Vert^2\\
			%             & V_R((A^a x^a + B^a u^a + E^a w), (A^a z^a + B^a v^a + E^a w), Sw) < V_R(x^a,z^a,w) - \Vert u - v\Vert^2\\
			%              &~~~+ c_R(\Vert u^a- v^a\Vert^2 + \Vert (C^a x^a + Fw) - (C^a z^a + Fw)\Vert^2), 
			%         \end{align}
		%     \end{subequations}
	%     with $x^a = (x,\zeta)$, $z^a = (z,\zeta)$, $u = E_2^\top \zeta + u^a$ and $v = E_2^\top \zeta + v^a$.
	% \end{assumption}
We can now summarize the closed-loop consequence for the implementable nominal DDPC scheme.

\begin{corollary}\label{cor:stability-aug}
	Suppose that the plant \eqref{eq:sys} satisfies Assumptions \ref{as:reg-eq}--\ref{as:nonresonance-T} and the data are persistently exciting of order $L+\iota$ in the sense of Definition~\ref{def:pe}.
	For any $\bar{Y}>0$, there exists $L_{\bar{Y}}>0$ such that, for all $L>L_{\bar{Y}}$ and all initial conditions $(\xi_0,\zeta_0)=\xi_0^a\in\X^a_{\bar{Y}}:=\{x^a\in\X^a~|~J_L^a(\xi^a)+W(\xi^a)\le\bar{Y}\}$, the DDPC problem \eqref{eq:ddmpc-incre} is recursively feasible and the augmented regulator manifold $\mathcal{A}^a:=\{\xi^a~|~\xi^a=\Pi^a\omega\}$ is exponentially stable for the closed loop.
\end{corollary}
\begin{proof}
	Lemma \ref{lem:aug-mpc-equ-ori-mpc} identifies \eqref{eq:ddmpc-incre} with the DDPC problem \eqref{eq:ddmpc-ori} posed on the augmented plant, and Proposition \ref{prop:aug-ctrl-detect-equ-ori-ctrl-detect} verifies the required structural assumptions for this augmented plant.
	Thus, the closed-loop properties follow from Theorem \ref{thm:stab-ori-ddmpc}.
\end{proof}

% Since Lemma \ref{lem:aug-mpc-equ-ori-mpc} shows that the incremental-input DDPC in \eqref{eq:ddmpc-incre} is equivalent to the feedforward-regularized nominal DDPC in \eqref{eq:ddmpc-ori} on the augmented plant, \eqref{eq:ddmpc-incre} is referred to as the nominal DDPC in the following sections.
Hereafter, we refer to the implementable incremental-input formulation
\eqref{eq:ddmpc-incre} as the nominal DDPC.

\begin{remark}[Increment-based predictors]
	\label{rem:increment-predictor}
	Under Assumption~\ref{as:period_exosignal}, an alternative predictor can be built from the period increments $\Delta_T u_t:=u_t-u_{t-T}$ and $\Delta_T y_t:=y_t-y_{t-T}$. Since $w_t-w_{t-T}=0$, these increments obey the disturbance-free dynamics $\Delta_T x_{t+1}=A\Delta_T x_t+B\Delta_T u_t$, $\Delta_T y_t=C\Delta_T x_t$, so that the standard fundamental lemma applies to them, and the output prediction can be recovered from $\bar y_{k|t}=\bar y_{k-T|t}+\Delta_T\bar y_{k|t}$, initialized with the measured outputs of the last period; see also \cite{dinkla2024data} for a predictive control scheme tailored to repetitive tasks. With noise-free data, both this increment-based predictor and \eqref{eq:ddmpc-incre:hankel} parametrize the exact trajectories of \eqref{eq:sys}; hence, combined with the cost \eqref{eq:ddmpc-incre:obj}, they lead to the same closed loop. The two predictors use different information. The increment-based predictor requires only the period $T$ and no phase information, but it needs $\iota+T$ past samples online and loses $T$ samples of the offline data. The predictor \eqref{eq:ddmpc-incre:hankel} needs only $\iota$ past samples, at the price of the phase information discussed in Remark~\ref{rem:phase-sync}. The robust analysis in Section~\ref{sec:ddmpc-noise} is developed for \eqref{eq:ddmpc-incre:hankel}.
\end{remark}

\section{Robust DDPC for Output Regulation\\ with Periodic Exosignals}
\label{sec:ddmpc-noise}

Throughout this section, Assumptions~\ref{as:period_exosignal} and \ref{as:nonresonance-T} are in
force, so the exosignal and the associated steady-state input are
$T$-periodic. We consider bounded additive noise in both the offline
output data and the online output measurements. These two noise sources
perturb the Hankel predictor and the initial-condition constraint,
respectively. Consequently, \eqref{eq:ddmpc-incre:hankel} no longer
parametrizes exact trajectories, and the nominal DDPC problem may become
infeasible or yield biased predictions. We therefore introduce slack and
coefficient regularization into \eqref{eq:ddmpc-incre} and establish
recursive feasibility and practical stability for the resulting closed loop.
\color{black}

\subsection{Robust DDPC}
\label{sec:ddmpc-noise:mpc-scheme}

Assume that both the output data used in the Hankel matrices and the online output measurements are corrupted by bounded additive noise:
$\tilde{y}_t=y_t+\varepsilon_t$, with $\|\varepsilon_t\|_\infty\le\bar{\varepsilon}$ for all $t\ge0$.
Given the noisy data $\{u_t,\tilde{y}_t\}_{t=0}^{N-1}$ and the recent measurements $(u_{[t-\iota,t-1]},\tilde{y}_{[t-\iota,t-1]})$, define the perturbed auxiliary state
\begin{equation}\label{eq:def-tilde-xi}
	\setlength{\abovedisplayskip}{3pt}
	\setlength{\belowdisplayskip}{3pt}
	\tilde{\xi}_t := \smat{\tilde{y}_{[t - \iota,t - 1]}\\
		u_{[t - \iota, t - 1]}} = \xi_t + \smat{\varepsilon_{[t - \iota, t - 1]}\\
		0}
\end{equation}
The robust DDPC problem is obtained by relaxing the output part of the Hankel equality and regularizing the coefficient vector
\begin{subequations}\label{eq:ddmpc-incre-noise}
	\setlength{\abovedisplayskip}{3pt}
	\setlength{\belowdisplayskip}{3pt} 
	\begin{align}
		\min_{\bar{y},\bar{u},g,h}~~& J^a_L(\tilde{\xi}_t, \zeta_t)= \sum_{k = 0}^{L- 1} \Vert \bar{y}_{k|t}\Vert^2_{Q} + \Vert \Delta\bar{u}_{k|t}\Vert^2_{R} \nonumber \\
		&~~+ \lambda_g \bar{\varepsilon}\Vert g_t - g^{\rm c}_{t}\Vert^2 + \frac{\lambda_h}{\bar{\varepsilon}}\Vert h_t\Vert^2  \label{eq:ddmpc-incre-noise:obj} \\
		{\rm s.t.}~~&\begin{bmatrix}
			\begin{smallmatrix}
				H_{L + \iota}(\tilde{y})\\
				H_{L + \iota}(u)\\
				H_1(\mathcal{M})
			\end{smallmatrix}
		\end{bmatrix} g_t = \begin{bmatrix}
			\begin{smallmatrix}
				\bar{y}_{[-\iota,L - 1]|t} + h_t\\
				\bar{u}_{[-\iota,L - 1]|t}\\
				\bar{M}_{t - \iota}
			\end{smallmatrix}
		\end{bmatrix}  \label{eq:ddmpc-incre-noise:hankel}\\
		& \begin{bmatrix}
			\begin{smallmatrix}
				\bar{y}_{[-\iota,-1]}(t)\\
				\bar{u}_{[-\iota,-1]}(t)
			\end{smallmatrix}
		\end{bmatrix} = \begin{bmatrix}
			\begin{smallmatrix}
				\tilde{y}_{[t - \iota, t - 1]}\\
				u_{[t - \iota,t - 1]}
			\end{smallmatrix}
		\end{bmatrix}  \label{eq:ddmpc-incre-noise:init-cond}\\
		& \bar{u}_{k|t} = \bar{u}_{k - T|t} + \Delta\bar{u}_{k|t},~~~k=0,\cdots,L-1  \label{eq:ddmpc-incre-noise:delta-u}\\
		& \bar{u}_{k - T|t} = \begin{cases}
			u_{t + k - T},&~~k < T\\
			\bar{u}_{k - T|t}, &~~k \ge T
		\end{cases}
		\label{eq:ddmpc-incre-noise:period-cond-u}\\
		& \bar{u}_{k|t} \in \mathbb{U},~~~k=0,\cdots,L-1\label{eq:ddmpc-incre-noise:bound-cond-u}.
	\end{align}  
\end{subequations}
% Here $g_t^{\rm c}$ is an available regularization center. In implementation, $g_t^{\rm c}=0$ is used during the first period, and $g_t^{\rm c}=g^*_{t-T}$ is used once $g^*_{t-T}$ is available.
% The regularizer $\lambda_g\bar{\varepsilon}\|g_t-g_t^{\rm c}\|^2$ limits amplification of noisy Hankel columns. The center $g_t^{\rm c}$ is periodic rather than zero because the exosystem induces a nonzero periodic steady-state coefficient sequence.
Here, $g_t^{\rm c}$ is a regularization center, and the regularizer $\lambda_g\bar{\varepsilon}\|g_t-g_t^{\rm c}\|^2$ limits amplification of noisy Hankel columns.
Since the regulated steady state is a nonzero periodic trajectory represented by a nonzero coefficient vector, the center cannot be chosen as zero.
We therefore compute the center offline from the data. Let $P_T$ be the matrix such that $P_T\bar{u}_{[-\iota,L-1]|t}$ stacks the increments $\bar{u}_{k|t}-\bar{u}_{k-T|t}$, $k=T-\iota,\ldots,L-1$ (the constraint involving $P_T$ is void if $L+\iota\le T$). For any exosystem basis vector $\bar{M}\in\mathbb{R}^{n_M}$ and a fixed weight $\mu>0$, define
\begin{equation}\label{eq:center-offline}
	\setlength{\abovedisplayskip}{3pt}
	\setlength{\belowdisplayskip}{3pt}
	\begin{aligned}
		g^{\rm c}(\bar{M}) := \arg\min_{g}~& \|H_{L+\iota}(\tilde{y})g\|^2+\mu\|g\|^2\\
		{\rm s.t.}~& \Psi g = \smat{\bar{M}\\ 0},~ \Psi:=\smat{H_1(\mathcal{M})\\ P_T H_{L+\iota}(u)},
	\end{aligned}
\end{equation}
and set $g_t^{\rm c}:=g^{\rm c}(\bar{M}_{t-\iota})$.
Problem \eqref{eq:center-offline} seeks a coefficient vector whose predicted output is approximately zero, whose predicted input is $T$-periodic, and whose exosystem basis vector is $\bar{M}$. It is thus a data-based approximation of the regulated steady state that requires neither $\Gamma_\xi$ nor the exosignal $w$.
Since the data are persistently exciting of order $L+\iota$, the matrix $\smat{H_1(\mathcal{M})\\ H_{L+\iota}(u)}$ has full row rank, and so does $\Psi$. Hence \eqref{eq:center-offline} is feasible for every $\bar{M}$, and its unique minimizer is linear in $\bar{M}$:
\begin{equation}\label{eq:center-closed-form}
	\setlength{\abovedisplayskip}{3pt}
	\setlength{\belowdisplayskip}{3pt}
	g_t^{\rm c}=G_{\rm c}\bar{M}_{t-\iota},~~
	G_{\rm c}:=\Phi^{-1}\Psi^\top\big(\Psi\Phi^{-1}\Psi^\top\big)^{-1}\smat{I_{n_M}\\ 0},
\end{equation}
where $\Phi:=H_{L+\iota}(\tilde{y})^\top H_{L+\iota}(\tilde{y})+\mu I\succ0$.
Thus, $G_{\rm c}$ is computed once offline, and only a matrix-vector product is needed online. Moreover, under Assumption~\ref{as:period_exosignal}, the basis $\mathcal{M}(\cdot)$ is $T$-periodic, so $g_t^{\rm c}$ is exactly $T$-periodic and does not depend on the closed-loop optimizers. This is in contrast with the recursive choice $g_t^{\rm c}=g^*_{t-T}$: the a priori bound on the optimizer only gives $\|g_t^*-g_t^{\rm c}\|^2\le\bar{Y}/(\lambda_g\bar{\varepsilon})$, cf. Appendix \ref{app:proof-lem4}, which deteriorates as $\bar{\varepsilon}\to0$ and may accumulate from one period to the next; hence, the recursive choice does not guarantee a uniformly bounded mismatch with the steady-state coefficient.

Compared with \eqref{eq:ddmpc-incre:hankel}, constraint \eqref{eq:ddmpc-incre-noise:hankel} contains slack variable $h_t$ that absorbs the mismatch caused by noisy output data. 
The penalty $(\lambda_h/\bar{\varepsilon})\|h_t\|^2$ discourages slack beyond what is justified by the noise level.  
With the chosen scaling, the regularizers vanish as $\bar{\varepsilon}\to0$, so \eqref{eq:ddmpc-incre-noise} recovers the nominal formulation. For fixed $\bar{\varepsilon}>0$, the problem is a convex quadratic program with linear constraints.

When \eqref{eq:ddmpc-incre-noise} is feasible at time $t$, let
$J_L^{a,*}(\tilde{\xi}_t,\zeta_t)$ denote its optimal value and
let $(\bar{y}_t^*,\bar{u}_t^*,g_t^*,h_t^*)$ denote an optimizer.
Analogously to the nominal scheme in Algorithm
\ref{alg:ddmpc-ori}, the corresponding robust receding-horizon
implementation is summarized in Algorithm
\ref{alg:ddmpc-incre-noise}. Its closed-loop properties are
established below.

\begin{algorithm}[h]
	\small
	\caption{Robust DDPC for Output Regulation.}
	\label{alg:ddmpc-incre-noise}
	\begin{algorithmic}[1]
		\State {\bfseries Given:} Prediction horizon $L \ge  \iota + {n}$, I/O data $\{u_t,\tilde{y}_t\}_{t = 0}^{N - 1}$ that are persistently exciting of order $L + \iota$ in the sense of Definition \ref{def:pe}, the constraint set $\mathbb{U}$, matrices $Q$, $R$, noise bound $\bar{\varepsilon}>0$,  regularization weights $\lambda_g > 0$, $\lambda_{h} > 0$, and exosignal period $T$, and the center map $G_{\rm c}$ computed offline from \eqref{eq:center-closed-form}.
		
		\State {\bfseries If} $t \le \iota - 1$, do \label{alg:ddmpc-noise-step1}				
		\State \quad Run \eqref{eq:sys} in open loop with randomly chosen input $u_t$, and store measurement $u_t$, $\tilde{y}_t$.
		
		\State {\bfseries Else if} $t \ge \iota$, do 
		\State  \quad Take the past $\iota$ I/O measurements $(u_{[t - \iota, t - 1]}, \tilde{y}_{[t - \iota, t - 1]})$, set $\bar{M}_{t - \iota}=\mathcal{M}(t-\iota)$ (cf. Remark~\ref{rem:phase-sync}), set $g_t^{\rm c}=G_{\rm c}\bar{M}_{t-\iota}$, and solve \eqref{eq:ddmpc-incre-noise}.
		
		\State \quad Apply the input $u_t = \bar{u}^*_{0|t}$.
		
		\State \quad  Set $t := t + 1$ and go back to \ref{alg:ddmpc-noise-step1}.
	\end{algorithmic}
\end{algorithm}

For the proof, it is useful to compare \eqref{eq:ddmpc-incre-noise} with a feedforward version of the same problem, paralleling the nominal analysis.
Define $g^s_t$ as a coefficient vector corresponding to the regulated steady-state trajectory, satisfying
\begin{subequations}\label{eq:hg=w-ss}
	\setlength{\abovedisplayskip}{3pt}
	\setlength{\belowdisplayskip}{3pt} 
	\begin{align}
		& \begin{bmatrix}
			\begin{smallmatrix}
				H_{L + \iota}({y})\\
				H_{L + \iota}(u)\\
				H_1(\mathcal{M})
			\end{smallmatrix}
		\end{bmatrix} g^s_t = \begin{bmatrix}
			\begin{smallmatrix}
				0_{p(L + \iota) \times 1}\\
				\Gamma_\xi \smat{
					S^{-\iota}\\
					\vdots\\
					I
				} \bar{w}_{[- \iota,  L- 1]|t}\\
				\bar{M}_{t - \iota}
			\end{smallmatrix}
		\end{bmatrix}, \\
		& \bar{w}_{k + 1|t} = S \bar{w}_{k|t}, ~~~ k = -\iota, \cdots, L - 2,\\
		& \bar{w}_{-\iota|t} = w_{t - \iota},
	\end{align}
\end{subequations}
whose minimum-norm solution is denoted by $g^{s*}_t$.
The sequence $g^{s*}_t$ is bounded for all $t \in \mathbb{N}$.
Under measurement noise, however, only the corrupted output $\tilde{y}$ is available.
We therefore use a noisy approximation of $g^{s*}_t$ obtained from
\begin{subequations}\label{eq:hg=w-ss-noise}
	\setlength{\abovedisplayskip}{3pt}
	\setlength{\belowdisplayskip}{3pt} 
	\begin{align}
		J_t^s &:= \min_{g^s, h^s}  \lambda_{g^s}\| g^s_t\|^2 + \lambda_{h^s} \| h^s_t\|^2 \\
		{\rm s.t.} ~~~& \begin{bmatrix}
			\begin{smallmatrix}
				H_{L + \iota}(\tilde{y})\\
				H_{L + \iota}(u)\\
				H_1(\mathcal{M})
			\end{smallmatrix}
		\end{bmatrix} g^s_t = \begin{bmatrix}
			\begin{smallmatrix}
				h^s_{[-\iota, L - 1]|t}\\
				\Gamma_\xi \smat{
					S^{-\iota}\\
					\vdots\\
					I
				} \bar{w}_{[- \iota,  L- 1]|t}\\
				\bar{M}_{t - \iota}
			\end{smallmatrix}
		\end{bmatrix} \\
		& \bar{w}_{k + 1|t} = S \bar{w}_{k|t}, ~~~ k = -\iota, \cdots, L - 2\\
		& \bar{w}_{-\iota|t} = w_{t - \iota}
	\end{align}
\end{subequations}
whose optimizer is denoted by $(\tilde{g}_t^s, \tilde{h}_t^s)$ and whose optimal cost is denoted by $J^{s*}_t$.
Then, assume that there exists $\bar{J}^s > 0$ such that $J^{s*}_t \le \bar{J}^s$ for all $t \in \mathbb{N}$.
This can be proved by using $\bar{g}^s_t := g^{s*}_t$, and $\bar{h}^s_t := H_{L + \iota}(\varepsilon) \bar{g}^{s*}_t$ as candidate solutions to \eqref{eq:hg=w-ss-noise}.
In addition, as in \cite{berberich2021linearII}, assume that there exists constant $c_{g^s} > 0$ such that $\|g^{s*}_t - \tilde{g}^s_t \|^2 \le c_{g^s}$; this bound is used in the stability analysis below.

% \begin{assumption}[Bounded mismatch]
% 	\label{as:robust-center}
% 	The regularization center $g_t^{\rm c}$ in \eqref{eq:ddmpc-incre-noise} is initialized such that there exists $c_{\rm c}>0$ satisfying $\|g_t^{\rm c}-\tilde g_t^s\|^2\le c_{\rm c}$ for all $t \ge 0$.
% \end{assumption}
%
% Assumption~\ref{as:robust-center} relates the implementable center
% $g_t^{\rm c}$ to the analytical reference $\tilde g_t^s$. The latter is
% used only in the proof to bound the coefficient regularization term; it is
% neither computed online nor used in Algorithm~\ref{alg:ddmpc-incre-noise}.

The analytical reference $\tilde g_t^s$ is used only in the proof to bound the coefficient regularization term; it is neither computed online nor used in Algorithm~\ref{alg:ddmpc-incre-noise}. The next result shows that the offline center \eqref{eq:center-offline} stays uniformly close to this reference, so that no additional assumption on the regularization center is needed.
\begin{proposition}[Bounded center mismatch]
	\label{prop:robust-center}
	Let Assumption~\ref{as:period_exosignal} hold, let the I/O data be persistently exciting of order $L+\iota$ in the sense of Definition~\ref{def:pe}, and let $\bar{\varepsilon}\in(0,\varepsilon_{\max}]$ for some $\varepsilon_{\max}>0$. Then the center $g_t^{\rm c}=G_{\rm c}\bar{M}_{t-\iota}$ in \eqref{eq:center-closed-form} satisfies
	\begin{equation}\label{eq:center-mismatch}
		\setlength{\abovedisplayskip}{3pt}
		\setlength{\belowdisplayskip}{3pt}
		\|g_t^{\rm c}-\tilde g_t^s\|^2\le c_{\rm c},\qquad \forall t\ge \iota,
	\end{equation}
	where $c_{\rm c}:=2\big(1+c_H^2/\mu\big)\|\Psi^\dagger\|^2c_M^2+2\bar{J}^s/\lambda_{g^s}$, $c_M:=\max_{0\le k\le T-1}\|\mathcal{M}(k)\|$, and $c_H:=\|H_{L+\iota}(y)\|+\sqrt{p(L+\iota)(N-L-\iota+1)}\,\varepsilon_{\max}$. In particular, $c_{\rm c}$ is independent of $t$ and of $\bar{\varepsilon}\in(0,\varepsilon_{\max}]$.
\end{proposition}
\begin{proof}
	Since every entry of $H_{L+\iota}(\varepsilon)$ is bounded by $\bar{\varepsilon}$ in magnitude, $\|H_{L+\iota}(\tilde y)\|\le\|H_{L+\iota}(y)\|+\|H_{L+\iota}(\varepsilon)\|_F\le c_H$. Under Assumption~\ref{as:period_exosignal}, $\mathcal{M}(\cdot)$ is $T$-periodic, so that $\|\bar{M}_{t-\iota}\|\le c_M$. The vector $g_0:=\Psi^\dagger\smat{\bar{M}_{t-\iota}\\0}$ is feasible for \eqref{eq:center-offline} because $\Psi$ has full row rank. Optimality of $g_t^{\rm c}$ gives
	\begin{equation*}
		\setlength{\abovedisplayskip}{3pt}
		\setlength{\belowdisplayskip}{3pt}
		\mu\|g_t^{\rm c}\|^2\le\|H_{L+\iota}(\tilde y)g_0\|^2+\mu\|g_0\|^2\le(c_H^2+\mu)\|\Psi^\dagger\|^2c_M^2.
	\end{equation*}
	Moreover, $J_t^{s*}\le\bar{J}^s$ yields $\|\tilde g_t^s\|^2\le\bar{J}^s/\lambda_{g^s}$, where, by the candidate solution given below \eqref{eq:hg=w-ss-noise}, $\bar{J}^s$ can be chosen uniformly for $\bar{\varepsilon}\in(0,\varepsilon_{\max}]$. The claim follows from $\|a-b\|^2\le2\|a\|^2+2\|b\|^2$.
\end{proof}

Proposition~\ref{prop:robust-center} guarantees the mismatch bound by construction; it is not a condition to be verified along the closed loop. The constant $c_{\rm c}$ is a worst-case bound. It enters the robust analysis only through regularization terms that are multiplied by $\lambda_g\bar{\varepsilon}$, cf. Appendix \ref{app:proof-lem4}, and therefore does not affect the vanishing of the ultimate bound as $\bar{\varepsilon}\to0$; a center closer to $\tilde g_t^s$ only reduces the corresponding constants.

With these preliminaries in place, 
recalling $(\Pi^a, \Gamma^a)$ denote the regulator pair of the augmented system in Proposition \ref{prop:aug-ctrl-detect-equ-ori-ctrl-detect},
we now introduce the following
feedforward-form representation of \eqref{eq:ddmpc-incre-noise} on the
augmented plant \eqref{eq:sys:aug}. This auxiliary formulation retains the
noisy Hankel model, the slack variable, and the implementable regularization
center $g_t^{\rm c}$, while expressing the incremental-input penalty through
the augmented input $u^a$:
\setlength{\abovedisplayskip}{3pt}
\setlength{\belowdisplayskip}{3pt}
\begin{subequations}
	\label{eq:ddmpc-aug-noise}
	\begin{align}
		\underset{\bar y,\bar u,\bar u^a,\atop
			\bar\zeta,\bar\omega,g,h}{\min}
		&J_L^{f}(\tilde\xi_t,\zeta_t,\omega_t)
		=\!\sum_{k=0}^{L-1}
		\Big(
		\|\bar y_{k|t}\|_Q^2
		\!+\|\bar u^a_{k|t}-\!\Gamma^a\bar\omega_{k|t}\|_R^2
		\Big) \notag\\
		&\quad+\lambda_g\bar\varepsilon\|g_t-g_t^{\rm c}\|^2
		+\frac{\lambda_h}{\bar\varepsilon}\|h_t\|^2
		\label{eq:ddmpc-aug-noise:obj}\\
		{\rm s.t.}\quad&
		\smat{
			H_{L+\iota}(\tilde y)\\
			H_{L+\iota}(u)\\
			H_1(\mathcal M)}
		g_t
		=
		\smat{
			\bar y_{[-\iota,L-1]|t}+h_t\\
			\bar u_{[-\iota,L-1]|t}\\
			\bar M_{t-\iota}},
		\label{eq:ddmpc-aug-noise:hankel}\\
		&
		\smat{\bar y_{[-\iota,-1]|t}\\
			\bar u_{[-\iota,-1]|t}}
		=
		\smat{\tilde y_{[t-\iota,t-1]}\\
			u_{[t-\iota,t-1]}},
		\label{eq:ddmpc-aug-noise:init}\\
		&\bar\zeta_{0|t}=\zeta_t,\label{eq:ddmpc-aug-noise:zeta-init}\\
		&\bar\zeta_{k+1|t}
		=E_0\bar\zeta_{k|t}\!+E_1\bar u^a_{k|t},
		~ k=0,\ldots,L\!-1,\label{eq:ddmpc-aug-noise:zeta}\\
		&\bar u_{k|t}
		=E_2^\top\bar\zeta_{k|t}+\bar u^a_{k|t},
		~~ k=0,\ldots,L-1,\label{eq:ddmpc-aug-noise:input}\\
		&\bar\omega_{0|t}
		=\smat{w_{t-\iota}\\ \vdots\\ w_t},~~
		\bar\omega_{k+1|t}=\mathbf S\bar\omega_{k|t},\\
		&\bar u_{k|t}\in\mathbb U,\quad k=0,\ldots,L-1.
	\end{align}
\end{subequations}
\begin{lemma}[Robust augmented reformulation]
	\label{lem:aug-mpc-equ-robust-mpc}
	Let Assumptions~\ref{as:reg-eq}--\ref{as:period_exosignal} hold.
	For fixed $t$, $\zeta_t$, and $\omega_t$, Problems
	\eqref{eq:ddmpc-incre-noise} and \eqref{eq:ddmpc-aug-noise}
	are equivalent.
%	 in the sense that their feasible sets are related
%	by an objective-preserving bijection.
\end{lemma}

\begin{proof}
%	See \ref{app:proof-lem3}.
%	See \ref{app:proof-lem3}.
	The equivalence follows from the same input-memory transformation as in
	\cite[Lemma~1]{kohler2021mpc-output-reg}
	(cf. Lemma~\ref{lem:aug-mpc-equ-ori-mpc}).
	The noisy Hankel constraint, slack variable $h$, and regularization terms are
	identical in \eqref{eq:ddmpc-incre-noise} and  \eqref{eq:ddmpc-aug-noise}, while
	$\bar u^a_{k|t}=\Delta\bar u_{k|t}$ and $\Gamma^a=0$.
	Hence, the transformation preserves feasibility and the objective value.
\end{proof}
Lemma \ref{lem:aug-mpc-equ-robust-mpc} allows us to establish recursive
feasibility and practical stability for Problem \eqref{eq:ddmpc-aug-noise}. The same
properties then hold for the implementable formulation
\eqref{eq:ddmpc-incre-noise}. Proposition~\ref{prop:robust-center} is used only
when bounding the coefficient regularization in the robust analysis below.
\color{black}

\subsection{Main Results}
\label{sec:ddmpc-noise:tech-result}
% \liu{The following analysis parallels the nominal case in Section~\ref{sec:ddmpc:stab-analy} and the standard robust DDPC analysis in \cite{berberich2021robust}. 
	% 	We therefore focus on the additional estimates caused by noisy output data and the periodic coefficient regularization center.}

For simplicity, let $\tilde{\xi}_t^a:=\smat{\tilde{\xi}_t\\\zeta_t}$, $
\tilde{\varrho}_t^a:=(\tilde{\xi}_t^a,\omega_t)$, and  define $\sigma^a(\tilde{\varrho}_t^a):=\|\tilde{\xi}_t^a-\Pi^a\omega_t\|^2$.
Noting that the output measurement noise affects only the first block of $\tilde{\xi}_t^a$, it follows that
$\|\tilde{\xi}_t^a-\xi_t^a\|\le\sqrt{p\iota}\,\bar{\varepsilon}$.
The following property, namely incremental uniform input/output-to-state stability (i-UIOSS), cf. \cite{kohler2021mpc-output-reg}, is stated first for the nominal augmented
system and then extended to the perturbed state by continuity.
\begin{proposition}\label{prop:i-UIOSS}
	Under Assumptions \ref{as:reg-eq}--\ref{as:nonresonance-T}, the augmented
	system \eqref{eq:sys:aug} is i-UIOSS. In particular, there exist a matrix
	$P_{a,W}=P_{a,W}^\top\succ0$ and a constant $\epsilon_o>0$ such that, with $W^a(\varrho_t^a):=
	\|\xi_t^a-\Pi^a\omega_t\|_{P_{a,W}}^2$,
	the following inequalities hold along every trajectory of
	\eqref{eq:sys:aug}:
	\begin{subequations}\label{eq:iuioss-bounds-W}
		\begin{align}
			&W^a(\varrho_t^a)
			\ge\epsilon_o\|\xi_t^a-\Pi^a\omega_t\|^2,
			\label{eq:iuioss-lower-W}\\
			&W^a(\varrho_{t+1}^a)
			\le W^a(\varrho_t^a)
			+\|u_t^a-\Gamma^a\omega_t\|_R^2+\|y_t\|_Q^2\nonumber\\
			&\quad-\epsilon_o\|\xi_t^a-\Pi^a\omega_t\|^2.
			\label{eq:iuioss-decrease-W}
		\end{align}
	\end{subequations}
	Moreover, there exists a $\mathcal{K}_\infty$ function
	$\alpha_W$ such that
	\begin{align}\label{eq:iuioss-perturbed-W}
		W^a(\tilde{\varrho}_{t+1}^a)
		&\le W^a(\tilde{\varrho}_t^a)
		+\|u_t^a\|_R^2+\|y_t\|_Q^2\nonumber\\
		&\quad-\epsilon_o\sigma^a(\tilde{\varrho}_t^a)
		+\alpha_W(\bar{\varepsilon}).
	\end{align}
\end{proposition}
\begin{proof}
%	See \ref{app:proof-prop5}.
%	See \ref{app:proof-prop5} in the Supplementary File.
By Proposition~\ref{prop:aug-ctrl-detect-equ-ori-ctrl-detect},
$(A^a,C^a)$ is detectable; hence, the nominal augmented error system admits
a quadratic i-UIOSS Lyapunov function, cf.\
\cite[Proposition~8]{kohler2021mpc-output-reg}, which yields
\eqref{eq:iuioss-bounds-W} after scaling.
Since $\Gamma^a=0$ and
$\|\tilde\xi_t^a-\xi_t^a\|\le\sqrt{p\iota}\bar\varepsilon$,
quadratic continuity of $W^a$ and Young's inequality give
\eqref{eq:iuioss-perturbed-W} for some
$\alpha_W\in\mathcal K_\infty$.
\end{proof}
Denoting the optimal value of Problem \eqref{eq:ddmpc-aug-noise} by $J_L^{f,*}(\tilde{\xi}_t,\zeta_t,\omega_t)$, consider the following Lyapunov candidate
\begin{equation}\label{eq:def-Yf}
	Y_L^f(\tilde{\varrho}_t^a)
	:=J_L^{f,*}(\tilde{\xi}_t,\zeta_t,\omega_t)
	+W^a(\tilde{\varrho}_t^a).
\end{equation}
The next lemma is the main technical estimate for the noisy
case. It shows that the Lyapunov candidate \eqref{eq:def-Yf} retains a one-step decrease up to a perturbation term
that is of class $\mathcal{K}_{\infty}$ in the noise bound.

\begin{lemma}
	\label{lem:continue-prop-Y-noise}
	Assume that Problem \eqref{eq:ddmpc-aug-noise} is feasible at time $t$,
	the I/O data are persistently exciting of order $L+\iota$ in the sense of Definition~\ref{def:pe}, and $Y_{L}^f(\tilde{\varrho}_t^a) \le \bar{Y}$.
	Then Problem \eqref{eq:ddmpc-aug-noise} with horizon $L-1$ is feasible at time $t+1$, and
	\begin{equation}\label{eq:continue-prop-Y-noise}
		Y_{L - 1}^f(\tilde{\varrho}_{t + 1}^a) \le Y_L^f(\tilde{\varrho}_t^a) - \epsilon_o \sigma^a(\tilde{\varrho}_t^a) + \alpha_3(\bar{\varepsilon})
	\end{equation}
	where $\alpha_3 \in \mathcal{K}_\infty$ and $\epsilon_o>0$ is as in Proposition \ref{prop:i-UIOSS}.
\end{lemma}
\begin{proof}
	The proof is deferred to Appendix \ref{app:proof-lem4}.
%	The proof can be found in Appendix C in the Supplementary File. 
	A noise-dependent
	prediction-error bound and a shifted feasible candidate at time $t+1$ are
	combined with a continuity estimate for the i-UIOSS Lyapunov function to
	obtain \eqref{eq:continue-prop-Y-noise}. The perturbation term
	$\alpha_3(\bar{\varepsilon})$ vanishes as
	$\bar{\varepsilon}\to0$, which is essential for the subsequent practical
	stability analysis.
\end{proof}
% The proof is given in Appendix \ref{app:proof-lem4}. It first bounds the prediction error caused by noisy Hankel data and the slack variable in \eqref{eq:ddmpc-noise:hankel}--\eqref{eq:ddmpc-noise:init-cond}. It then constructs a shifted feasible candidate at time $t+1$ and combines the resulting value-function estimate with continuity of the i-UIOSS Lyapunov function. The term $\alpha_3(\bar{\varepsilon})$ tends to zero as $\bar{\varepsilon}\to0$, which is the mechanism behind practical stability.

The robust closed-loop guarantee can now be stated.

% \subsection{Closed-Loop Stability Guarantees}
% \label{sec:ddmpc-noise:stab}
\begin{theorem}
	[Robust stability guarantee]
	\label{thm:ddmpc-stab-noise}
	Under Assumptions \ref{as:reg-eq}--\ref{as:nonresonance-T}, suppose that the I/O data are persistently exciting of order $L+\iota$ in the sense of Definition~\ref{def:pe}, and let the regularization center be given by \eqref{eq:center-closed-form}.
	If $Y_L^f(\tilde{\varrho}_0^a)\le\bar{Y}$, then there exist $\tilde{L}_{\bar{Y}}\ge0$ and $\hat{\varepsilon}_L>0$ such that, for all $L>\tilde{L}_{\bar{Y}}$ and all  $\bar{\varepsilon}\le\hat{\varepsilon}_L$, the robust DDPC problem \eqref{eq:ddmpc-incre-noise} is feasible for all $t\ge0$, the closed-loop input satisfies $u_t\in\mathbb{U}$, and the Lyapunov candidate $Y_L^f=J_L^{f,*}+W^a$ satisfies
	\setlength{\abovedisplayskip}{3pt}
	\setlength{\belowdisplayskip}{3pt}
	\begin{subequations}\label{eq:up-low-bound-decrease-Y-noise}
		\begin{align}
			\epsilon_o \sigma^a(\tilde{\varrho}_t^a) \le Y_L^f(\tilde{\varrho}_t^a) \le \gamma_{\bar{Y}}\sigma^a(\tilde{\varrho}_t^a)  + \alpha_L(\bar{\varepsilon}), \label{eq:up-low-bound-Y-noise}\\
			Y_{L}^f(\tilde{\varrho}_{t + 1}^a) - Y_L^f(\tilde{\varrho}_t^a) \le -\tilde{\alpha}_L \sigma^a(\tilde{\varrho}_t^a)  + \alpha_8(\bar{\varepsilon}),\label{eq:decrease-Y-noise}
		\end{align}
	\end{subequations}
	with $\gamma_{\bar{Y}},\tilde{\alpha}_L>0$ and $\alpha_L,\alpha_8\in\mathcal{K}_{\infty}$. Consequently, the augmented regulator manifold is practically stable. Since
	$\|\tilde{\xi}_t-\Pi_\xi\omega_t\|^2\le
	\sigma^a(\tilde{\varrho}_t^a)$, the original regulation error is
	practically stable as well, with an ultimate bound that vanishes as
	$\bar{\varepsilon}\to0$.
\end{theorem}
\begin{proof}
	See Appendix \ref{app:proof-thm3}.
%	See Appendix D in the Supplementary File.
\end{proof}

\begin{remark}[Relation to existing robust DDPC]
	Although the stability argument in Theorem \ref{thm:ddmpc-stab-noise} follows the general robust DDPC strategy of \cite{berberich2021robust}, the coefficient regularization requires a different treatment because the regulated steady state is a nonzero periodic trajectory. In particular, the implementable center $g_t^c$ generally differs from the analytical steady-state coefficient $\tilde{g}_t^s$. The offline center \eqref{eq:center-offline} keeps this mismatch uniformly bounded by construction (cf. Proposition~\ref{prop:robust-center}), which allows the shifted-candidate argument to retain a vanishing noise-dependent perturbation. Theorem \ref{thm:ddmpc-stab-noise} therefore extends the robust DDPC guarantee from equilibrium stabilization to periodic output regulation.
\end{remark}

\section{Numerical Validation}
\label{sec:numerical_example}

We validate the proposed controllers on a constrained autonomous-vehicle lateral-control problem on a curvilinear road, adapted from \cite{snider2009automatic,zuliani2026policy}. The example is deliberately chosen as an output regulation task rather than a setpoint-tracking task: a periodic curvature signal acts as an exogenous disturbance, the regulated output is the lateral displacement error, and the steering input is constrained. The controller is designed from input-output data only. The state is $x=\mathrm{col}(e_{cg},\dot e_{cg},\theta_e,\dot\theta_e)$, where $e_{cg}$ is the lateral displacement error of the vehicle center of gravity and $\theta_e$ is the heading error. The steering angle is denoted by $u$. The continuous-time model used to generate data and closed-loop simulations is
$\dot x(t)=A_cx(t)+B_cu(t)+E_cr(t)$, where $r(t)=v_x\kappa(t)$ is induced by the path curvature $\kappa(t)$ and
\setlength{\abovedisplayskip}{3pt}
\setlength{\belowdisplayskip}{3pt}
\[
A_c= \smat{
	0&1&0&0\\
	0&a_1&a_2&a_3\\
	0&0&0&1\\
	0&a_4&a_5&a_6
},~~
B_c=
\smat{
	0\\ b_1\\ 0\\ b_2
},~~
E_c=
\smat{
	0\\ a_3-v_x\\ 0\\ a_6
}.
\]
The numerical parameters are $a_1=-27.280$, $a_2=272.798$, $a_3=a_4=a_5=0$, $a_6=-29.388$, $b_1=136.399$, $b_2=126.129$, and $v_x=10~\mathrm{m/s}$. The model is discretized by forward Euler with sampling time $T_s=0.05~\mathrm{s}$, giving $A=I+T_sA_c$, $B=T_sB_c$, and $E_r=T_sE_c$. The regulated output is $y_t=Cx_t$ with $C=[1~0~0~0]$,
so that the output regulation objective coincides with asymptotic lane-centering.

\begin{figure}
	\centering
	\includegraphics[width=1\columnwidth]{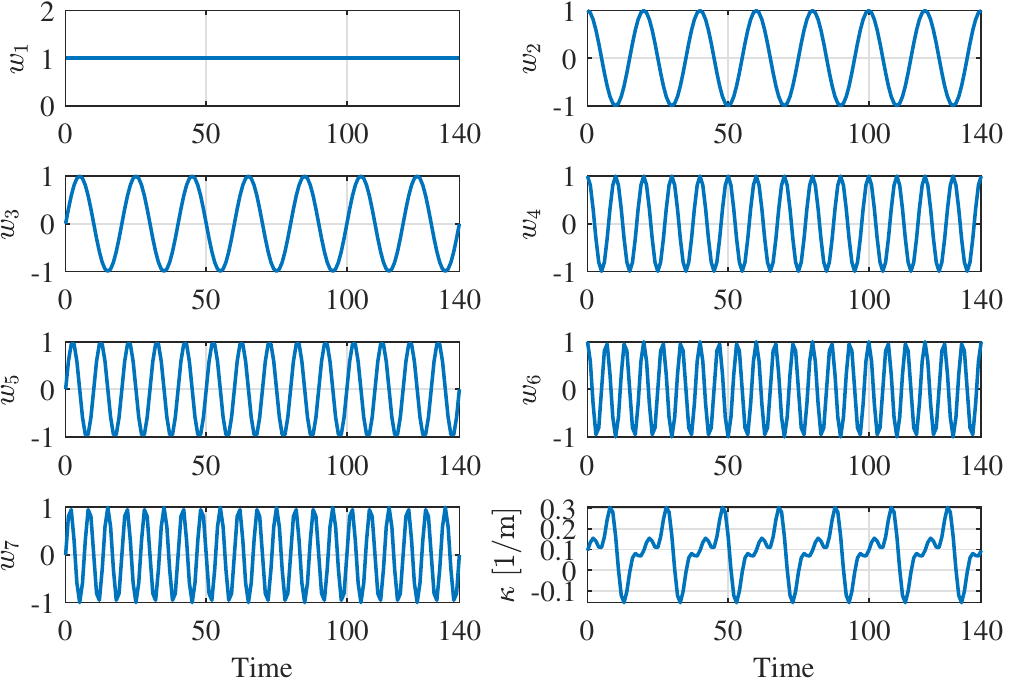}
	\caption{Periodic exogenous signal and induced road curvature. The signal contains one constant mode and three harmonic pairs with period $T_p=20$.}
	\label{fig:vehicle_exosignal}
\end{figure}

\begin{figure}
	\centering
	\includegraphics[width= 1\columnwidth]{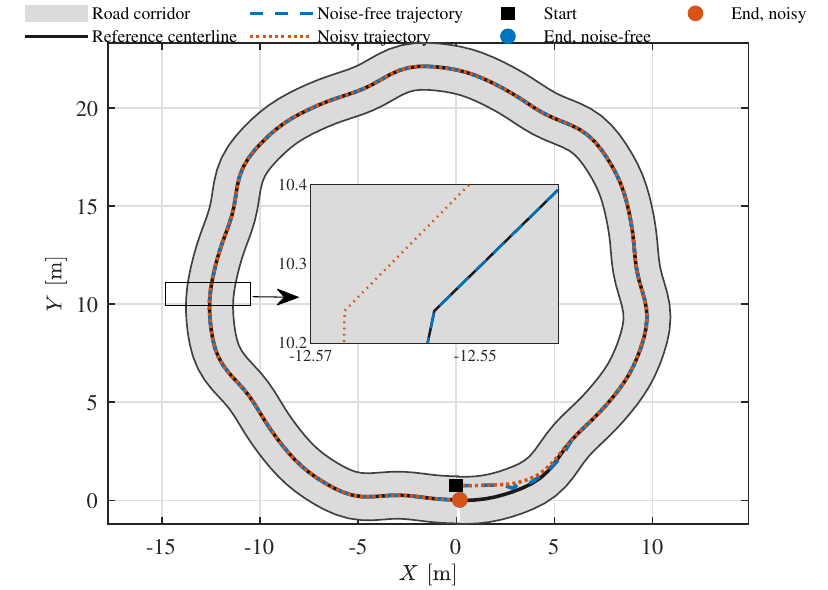}
	\caption{Reconstructed curvilinear road and closed-loop vehicle trajectories. The gray corridor is centered on the curvature-generated reference path; vehicle trajectories are obtained by offsetting the centerline by the simulated lateral error $e_{cg}$.}
	\label{fig:vehicle_path}
\end{figure}

The road curvature is generated by the seven-dimensional periodic exosystem
\begin{equation}
	w_{t+1}=Sw_t,\qquad
	S={\rm block.diag}(1,S_1,S_2,S_3),
	\label{eq:vehicle_exosystem}
\end{equation}
where $S_i=
\smat{
	\cos(i\omega)&-\sin(i\omega)\\
	\sin(i\omega)& \cos(i\omega)
}$ for $i=1,2,3$, 
%\[
%S_i=
%\smat{
%	\cos(i\omega)&-\sin(i\omega)\\
%	\sin(i\omega)& \cos(i\omega)
%},\qquad i=1,2,3,
%\]
$\omega=2\pi/T_p$, and $T_p=20$. The initial exogenous signal is $w_0=[1~~1~~0~~1~~0~~1~~0]^{\top}$.
Thus $w_t$ contains a constant component and three harmonic pairs. The curvature and yaw-rate input are
$ \kappa_t = [0.090~~0~~0.120~~0.040~~-0.090~~-0.035~~0.060]w_t$ and $r_t=G_w w_t$, where $G_w=v_x
[
0.090~~0~~0.120~~0.040~~-0.090~~-0.035~~0.060]$.
The resulting discrete-time system is
\setlength{\abovedisplayskip}{3pt}
\setlength{\belowdisplayskip}{3pt}
\begin{equation}
	x_{t+1}=Ax_t+Bu_t+Ew_t,\qquad y_t=Cx_t,
	\label{eq:exm_car}
\end{equation}
where $E=E_rG_w$. Fig.~\ref{fig:vehicle_exosignal} shows the exogenous modes and the induced curvature. For visualization only, the reference path in Fig.~\ref{fig:vehicle_path} is reconstructed by integrating the prescribed curvature, and the vehicle position is displayed by offsetting the centerline according to the simulated lateral error $e_{cg}$. 
The matrices in \eqref{eq:exm_car} are used only to generate data and simulate the true plant; they are not supplied to the DDPC optimizations. 


\begin{figure}
	\centering
	\includegraphics[width=1\columnwidth]{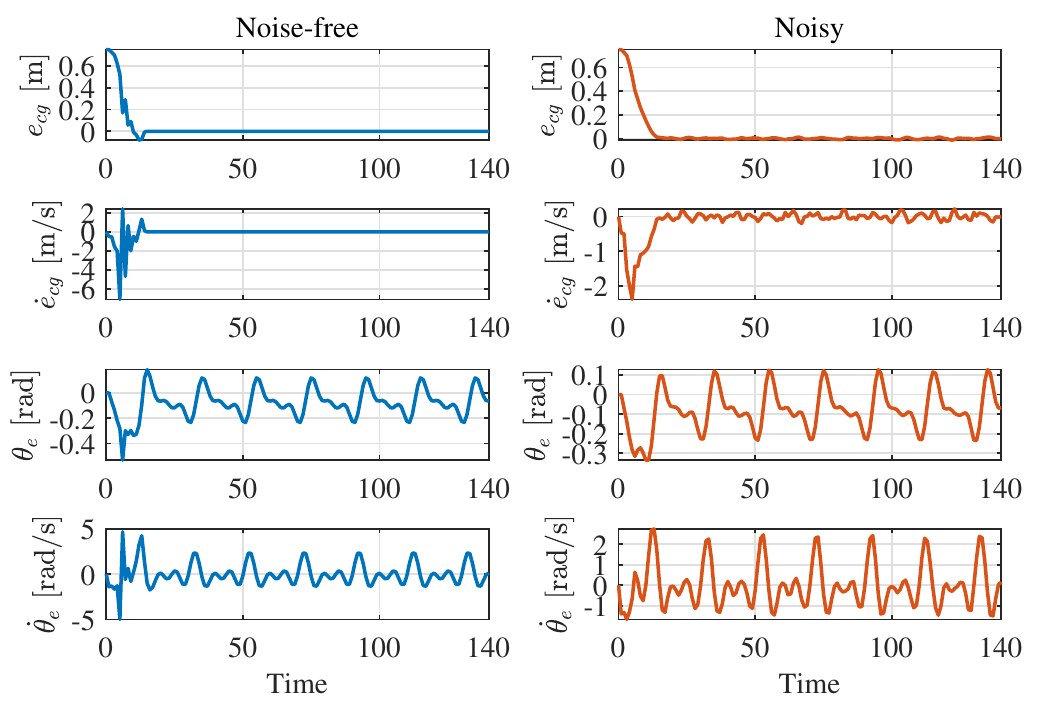}
	\caption{Closed-loop state trajectories for nominal DDPC with noise-free data (left column) and robust DDPC with noisy data and noisy online outputs (right column).}
	\label{fig:vehicle_states}
\end{figure}

\begin{figure}
	\centering
	\includegraphics[width=1\columnwidth]{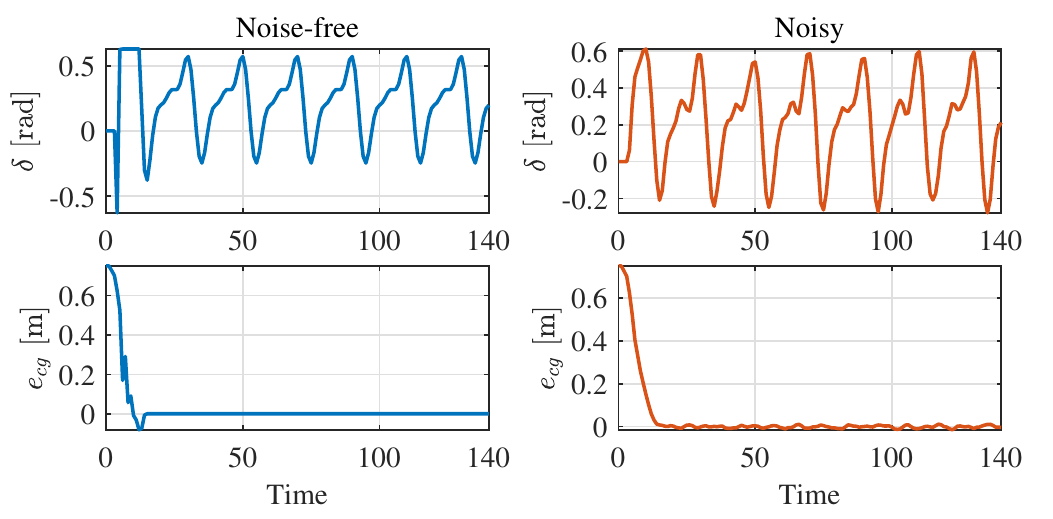}
	\caption{Closed-loop steering input and regulated lateral error. The steering constraint is respected in both the noise-free and noisy cases.}
	\label{fig:vehicle_input_output}
\end{figure}

The offline experiment consists of one input-output trajectory of length $N=800$. 
During data collection, the plant is excited by $u_t=Kx_t+\bar u_t$, where $\bar u_t$ is uniformly sampled and the steering input is saturated to $[-\pi/5,\pi/5]$. 
The stabilizing matrix $K$ is used only to collect informative data and is not used by the online controller. 
The observability index is $\iota=4$, and the prediction horizon is $L=40$. The invariant zeros of $(A,B,C)$ are $0.2653\pm0.3014{\rm i}$, with modulus $0.40$; hence, Assumption~\ref{as:nonresonance-T} holds. 
% Figs. \ref{fig:vehicle_states} and \ref{fig:vehicle_input_output} report two cases side by side: nominal DDPC with exact data and robust DDPC with corrupted offline and online outputs.

\begin{figure}
	\centering
	\includegraphics[width=1\columnwidth]{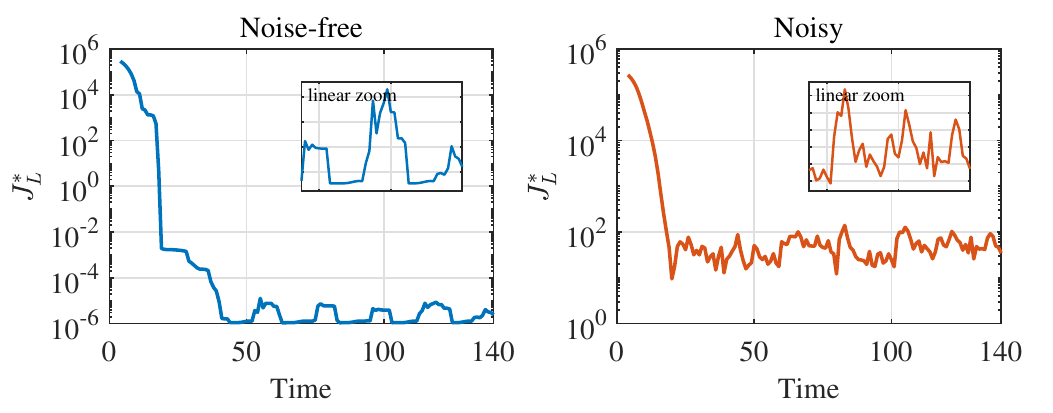}
	\caption{Optimal values returned by the nominal and robust DDPC optimization problems along the closed-loop trajectories.}
	\label{fig:vehicle_cost}
\end{figure}

We first test the nominal DDPC with periodic incremental-input regularization using noise-free data. The weights are $Q=10^5$ and $R=10^{-3}$, the initial condition is $x_0=[0.75~~0~~0~~0]^{\top}$, and the steering constraint is $u_t\in[-\pi/5,\pi/5]$. Figs.~\ref{fig:vehicle_states} and \ref{fig:vehicle_input_output} show that the lateral error is regulated while the steering input remains within the imposed bounds. Fig.~\ref{fig:vehicle_cost} shows the corresponding optimal values, illustrating the decrease of the receding-horizon objective along the closed-loop trajectory. The spatial reconstruction in Fig.~\ref{fig:vehicle_path} confirms that regulation of $e_{cg}$ translates into path-following performance in curvilinear coordinates.

We then repeat the experiment with noisy offline output data and noisy online output measurements. The additive output noise satisfies $|\varepsilon_t|\leq\bar\varepsilon$, with $\bar\varepsilon=5\times10^{-3}$. The robust DDPC formulation is used with $Q=10^5$, $R=10^{-3}$, and regularization weights $\lambda_g\bar{\varepsilon}=10^4$ and $\lambda_h/\bar{\varepsilon}=10^{15}$. 
The regularization center $g_t^{\rm c}$ is calculated offline based on \eqref{eq:center-offline}--\eqref{eq:center-closed-form} with $\mu = 10^{-4}$.
The same initial condition, horizon, and steering constraint as in the nominal test are used, so that the effect of measurement noise is isolated. As shown in Figs.~\ref{fig:vehicle_path}--\ref{fig:vehicle_input_output}, the robust controller preserves constraint satisfaction and keeps the regulated output in a small neighborhood of zero despite corrupted data and online measurements. This behavior is consistent with Theorem~\ref{thm:ddmpc-stab-noise}: exact convergence is replaced by practical regulation, and the residual error is governed by the measurement-noise level.

\section{Conclusions}
% This paper has developed a predictive-control approach to data-driven output regulation for unknown linear systems. The central idea is a trajectory parametrization for input-output data generated in the presence of dynamic exosignals, obtained by combining Willems et al's fundamental lemma with the known exosystem structure. This parametrization enables nominal receding-horizon regulators with exponential convergence to the regulator manifold. For periodic exosignals, the steady-state feedforward term can be removed from the online optimization and replaced by an incremental-input penalty, yielding an implementable data-driven regulator. A robust formulation with slack and coefficient regularization was shown to be recursively feasible and practically stabilizing under bounded output noise. The vehicle lateral-control study illustrates the main features of the theory: rejection of periodic curvature disturbances, steering-constraint satisfaction, and practical regulation when both offline and online outputs are noisy.
%
% Future work will address less structured exogenous signals, data-efficient excitation conditions, and distributed output regulation settings in which multiple agents share only local input-output measurements.

This paper has developed a predictive-control approach to data-driven output regulation for unknown linear systems. The central idea is a trajectory parametrization for input-output data generated by exosignals with known dynamics and known phase, obtained by combining Willems et al.'s fundamental lemma with the known exosystem structure. This parametrization enables constrained receding-horizon regulators with exponential convergence to the regulator manifold, without a plant model or the solution of the regulator equations. For periodic exosignals, the steady-state feedforward term is removed from the online optimization by an incremental-input penalty, yielding an implementable data-driven regulator. A robust formulation with slack and coefficient regularization, whose regularization center is computed offline from data, was shown to be recursively feasible and practically stabilizing under bounded noise in both the offline data and the online measurements. The vehicle lateral-control study illustrates the main features of the theory: rejection of periodic curvature disturbances, steering-constraint satisfaction, and practical regulation when both offline and online outputs are noisy.

Future work will address exosignals whose phase is not available, non-periodic exogenous signals, data-efficient excitation conditions, and distributed output regulation settings in which multiple agents share only local input-output measurements.

\bibliographystyle{IEEEtran}
\bibliography{ddPiMPC}

%\clearpage

%\noindent\textbf{\Large Supplementary File of Data-Driven Output Regulation via Predictive Control}
%
%   \noindent\textit{Equation, result, and reference numbers refer to the main manuscript.}
\appendices

\section{Proof of Theorem \ref{thm:stab-ori-ddmpc}}
\label{app:proof-thm2}
\begin{proof}
	%	Theorem \ref{thm:fund-lem-exo} ensures that the DDPC problem is equivalent to the model-based MPC in \cite[eq. (5)]{kohler2021mpc-output-reg}.
	%	Hence, under Propositions \ref{prop:stab} and \ref{prop:i-IOSS},  we can apply \cite[Theorem 1]{kohler2021mpc-output-reg} to conclude that the function $Y_L(\xi_t,\omega_t)$ satisfies 
	Theorem~\ref{thm:fund-lem-exo} ensures that the DDPC prediction model
	parametrizes the same finite-horizon input-output trajectories as the
	corresponding model-based formulation.
	Together with Propositions~\ref{prop:stab} and \ref{prop:i-IOSS}, this
	provides the local cost bound and i-IOSS estimate required by the
	finite-horizon Lyapunov argument. Consequently,
	$Y_L(\xi_t,\omega_t)$ satisfies 
	\begin{subequations}\label{eq:low-up-bound-decrease-Y}
		\begin{align}
			\epsilon_o \sigma(\varrho_t) & \le Y_L(\varrho_t) \le \gamma_{\bar{Y}} \sigma(\varrho_t),\label{eq:low-up-bound-Y}\\
			Y_L(\varrho_{t + 1}) &- Y_L(\varrho_t) \le -\alpha_L\epsilon_o \sigma(\varrho_t) \label{eq:decrease-Y},        \end{align}
	\end{subequations}
	for the resulting closed-loop with 
	%{	\setlength{\abovedisplayskip}{3pt}
		%	\setlength{\belowdisplayskip}{3pt}
		\begin{equation}\label{eq:alpha_N}
			\alpha_L := 1 - \frac{\gamma_s \gamma_{\bar{Y}}}{\epsilon_o^2(L - 1)} >0.
		\end{equation}
		Further, the closed-loop system satisfies the following transient performance bound
		%	{\setlength{\abovedisplayskip}{3pt}
			%	\setlength{\belowdisplayskip}{3pt}
			\begin{equation}\label{eq:relation-alpha-Y_N-barY}
				\alpha_L\sum_{t = 0}^{L - 1} \ell(\xi_t,\omega_t) \le Y_L(\varrho_0)\le \bar{Y},~~\forall L \in \mathbb{N}.
			\end{equation}
			The following proof is for completeness of the statements.
			We first show inequalities \eqref{eq:low-up-bound-Y}-\eqref{eq:decrease-Y} for all $\varrho_t \in \X^{\varrho}_{\bar{Y}}$, and then the performance bound \eqref{eq:relation-alpha-Y_N-barY}. Abbreviate $\ell_k := \ell(\varrho_{k|t}^*)$.

			\textbf{Step I.} The lower bound in \eqref{eq:low-up-bound-Y} follows with $\ell_0\ge 0$, and
			\begin{equation}
				Y_L(\varrho_t) \overset{\eqref{eq:ddmpc-ori:obj}}{\ge} \ell_0  + W(\varrho_t)\overset{\eqref{eq:iioss-lower-W}}{\ge} \epsilon_o \sigma(\varrho_t).
			\end{equation}
			Recall that $W(\varrho_t) = \Vert \xi_t - \Pi_\xi  \omega_t\|^2_{P_{\xi,W}} \le \lambda_{\max}(P_{\xi,W}) \sigma(\varrho_t)$ $:= \gamma_o \sigma(\varrho_t)$.
			For any $\varrho_t \in \X^{\varrho}_{\delta}$, we directly obtain the bound $Y_L(\varrho_t) \le (\gamma_s + \gamma_{o}) \sigma(\varrho_t)$ using \eqref{eq:up-bound-J}.
			In addition, for $\varrho_t \notin \X^{\varrho}_{\delta}$, it follows from $\varrho_t \in \X^{\varrho}_{\bar{Y}}$ and $\sigma (\varrho) \ge\delta_s$ that $Y_L(\varrho_t) \le \frac{\bar{Y}}{\delta_s} \sigma(\varrho_t)$.
			Hence, the upper bound in \eqref{eq:low-up-bound-decrease-Y} holds with $\gamma_{\bar{Y}} := \max\{\gamma_s + \gamma_o,  \frac{\bar{Y}}{\delta_s}\}$.
			%	, which is similar to \cite[Theorem 2]{kohler2018nonlinear-track-mpc}.
			
			\textbf{Step II.} The i-IOSS Lyapunov condition \eqref{eq:iioss-decrease-W} implies
			\begin{align}\label{eq:iterition-W}
				\setlength{\abovedisplayskip}{3pt}
				\setlength{\belowdisplayskip}{3pt} 
				W(\varrho_{L|t}^*) - W(\varrho_t) &= \sum_{k = 0}^{L - 1} W(\varrho^*_{k + 1| t}) - W(\varrho_{k|t}^*) \nonumber\\
				&~~\overset{\eqref{eq:iioss-decrease-W}}{\le} \sum_{k = 0}^{L - 1} - \epsilon_o \sigma(\varrho_{k|t}^*) + \sum_{k = 0}^{L - 1}\ell_k.
			\end{align}
			Using $W(\varrho_{L|t}^*) \ge 0$, and $\varrho_t \in \X^{\varrho}_{\bar{Y}}$, we arrive at 
			\begin{equation}
				\epsilon_o \sum_{k = 0}^{L - 1} \sigma(\varrho_{k |t}^*) \overset{\eqref{eq:iterition-W}}{\le }Y_L(\varrho_t) \overset{\eqref{eq:low-up-bound-Y}}{\le}\min\{\bar{Y}, \gamma_{\bar{Y}}\sigma(\varrho_t) \}.
			\end{equation}
			Thus, there exists integer $k_{\varrho} \in \{1,\cdots,L - 1 \}$, such that 
			\begin{equation}\label{eq:up-bound-sigma}
				\sigma(\varrho^*_{k_\varrho}|t) \le \frac{\min\{ \bar{Y}, \gamma_{\bar{Y}}\sigma(\varrho_t)\}}{\epsilon_o(L - 1)}.
			\end{equation}
			Given $L \ge L_0 := 1 + \frac{\bar{Y}}{\delta_s \epsilon_o}$, this implies $\varrho_{k_\varrho|t}^* \in \X^{\varrho}_{\delta}$.
			Thus, Proposition \ref{prop:stab} ensures that starting at $\varrho_{k_\varrho|t}^*$ there exists feasible state and input trajectory satisfying the bound  \eqref{eq:up-bound-J}, which implies 
			\begin{align}\label{eq:iteration-V}
				& J_{L}(\varrho_{t + 1}) + \ell_0 \le \sum_{k = 0}^{k_\varrho - 1}\ell_k + J_{L - k_\varrho + 1}(\varrho_{^*_{k_\varrho|t}}) \nonumber \\
				&\overset{\eqref{eq:up-bound-J}}{\le }J_L(\varrho_t) + \gamma_s \sigma(\varrho_{k_\varrho|t}^*) \overset{\eqref{eq:up-bound-sigma}}{\le} J_L(\varrho_t) + \frac{\gamma_s\gamma_{\bar{Y}}}{\epsilon_o(L - 1)} \sigma(\varrho_t).
			\end{align}
			Combining \eqref{eq:iteration-V} and \eqref{eq:iioss-decrease-W}, the function $Y_L$ satisfies \eqref{eq:decrease-Y}.
			Furthermore, $\alpha_L > 0$ follows from \eqref{eq:alpha_N}, using $L > L_1 := 1 + \gamma_s \gamma_{\bar{Y}}/\epsilon_o^2$.
			All the arguments hold with $L > L_{\bar{Y}} := \max\{L_0,L_1\}$.
			In addition, inequality \eqref{eq:decrease-Y} ensures $Y_L$ is nonincreasing and thus $\varrho_t \in \X_{\bar{Y}}$ for all $t \ge 0$.
			
			\textbf{Step III.} Using $Y_L = J_L + W \ge J_L \ge 0$, $\alpha_L \le 1$ and the following inequality in the telescopic sum
			\begin{align}\label{eq:iteration-YW}
				&Y_{L}(\varrho_{t + 1}) - Y_{L}(\varrho_t) - \alpha_L(W(A_\varrho \varrho + B_\varrho u) - W(\varrho)) \\
				&= (1 - \alpha_L)(Y_{L}(\varrho_{t + 1}) - Y_L(\varrho_t)) + \alpha_L(J_L(\varrho_{t + 1}) - J_L(\varrho_t)) \nonumber\\
				&\overset{\eqref{eq:relation-alpha-Y_N-barY},\eqref{eq:iteration-V}}{\le}-\alpha_L\ell(\varrho_t),
			\end{align}
			the performance bound \eqref{eq:relation-alpha-Y_N-barY} is established.
			
			Inequalities \eqref{eq:low-up-bound-Y}-\eqref{eq:decrease-Y} directly imply $\lim_{t \rightarrow \infty}\sigma(\varrho_t) = 0$ and hence asymptotic convergence.
			Building on this result, in the following, we show that $\mathcal{A}_\varrho$ is exponentially stable.
			Define the point-to-set distance $\Vert \varrho\Vert_{\mathcal{A}_\varrho} := \inf_{s \in \mathcal{A}_\varrho}\Vert \varrho - s \Vert$.
			In the following, we show that there exists constant $c_{\pi} > 0$, such that $c_\pi \sigma(\varrho) \le \Vert \varrho \Vert_{\mathcal{A}_\varrho}^2 \le \sigma(\varrho)$, which in combination with \eqref{eq:low-up-bound-decrease-Y}-\eqref{eq:relation-alpha-Y_N-barY} ensures exponential stability of $\mathcal{A}_\varrho$ using standard Lyapunov arguments. 
			For given $(\xi,\omega)$, denote some minimizer by $\tilde{\omega} := \arg\min_{\tilde{\omega} \in \mathbb{\Omega}} \Vert  (\Pi_\xi \tilde{\omega}, \tilde{\omega}) - (\xi,\omega)\Vert$. 
			We can deduce that 
			\begin{align*}
				\sigma(\varrho) & = \Vert \xi - \Pi_\xi \omega\Vert^2\\
				& \le 2 (\Vert \xi - \Pi_\xi \tilde{\omega}\Vert^2 + \Vert \Pi_\xi\tilde{\omega} - \Pi_\xi \omega \Vert^2)\\
				& \le 2\max\{1,\|\Pi_\xi\|^2\}\Vert (\xi,\omega) - (\Pi_\xi \tilde{\omega},\tilde{\omega})\Vert^2 := \frac{1}{c_\pi} \Vert \varrho \Vert_{\mathcal{A}_\varrho}^2
			\end{align*}
			where the first inequality uses $\Vert a + b\Vert^2 \le 2(\Vert a\Vert^2 + \Vert b\Vert^2)$ for any $a,b \in \R^{n}$.
			Further, 
			\begin{equation}
				\Vert \varrho \Vert_{\mathcal{A}_\varrho} = \Vert (\Pi_\xi \tilde{\omega},\tilde{\omega}) - (\xi,\omega) \Vert \le \Vert \xi - \Pi_\xi \omega\Vert \overset{\eqref{eq:sigma-varrho}}{=} \sqrt{\sigma(\varrho)}.
			\end{equation}
			Hence, $\lim_{t \rightarrow\infty} \Vert \varrho \Vert_{\mathcal{A}_\varrho} = 0$ as $\lim_{t \rightarrow \infty} \sigma(\varrho) = 0$, 
			which completes the proof.
		\end{proof}

\section{Proof of Proposition \ref{prop:aug-ctrl-detect-equ-ori-ctrl-detect}}
\label{app:proof-prop4}

\begin{proof}
	\emph{Regulator equations:}
	Since $(\Pi_\xi,\Gamma_\xi)$ solves the regulator equations \eqref{eq:aux-reg-eq}  and $\mathbf{S}^T=I$ under
	Assumption~\ref{as:period_exosignal}, the augmented regulator equations
	are solved by
	\begin{equation}
		\Pi^a = \smat{\Pi_\xi\\
			\Gamma_\xi \mathbf{S}^{T - 1}\\
			\Gamma_\xi \mathbf{S}^{T - 2}\\
			\vdots\\
			\Gamma_\xi},~~\Gamma^a = 0.
	\end{equation}
	
	% In the following, we show that the augmented system \eqref{eq:sys:aug} is stabilizable and detectable.
	
	\emph{Stabilizability and detectability:}
	Recalling that matrices $(A^a, B^a, C^a)$ of the augmented system \eqref{eq:sys:aug} are given by
	\begin{equation*}
		A^a = \smat{\mathbf{A}_\iota & \mathbf{B}_\iota E_2^\top\\
			0 & E_0},~~B^a = \smat{\mathbf{B}_\iota\\
			E_1},~~C^a = \smat{Z_1 & 0}.
	\end{equation*}
	Stabilizability of $(A^a,B^a)$ is equivalent to 
	\begin{equation}\label{eq:rank-ctrl-augsys}
		{\rm rank} \smat{\mathbf{A}_\iota  - \lambda I_{n_\xi} & \mathbf{B}_\iota E_2^\top & \mathbf{B}_\iota\\
			0 & E_0 - \lambda I_{mT} & E_1} = n_\xi + Tm,
	\end{equation}
	for all $\lambda\in\C$ satisfying $|\lambda|\ge1$.
	Consider $\lambda_k = e^{{\rm i}2\pi k/T}$ for $k = 0,\cdots, T - 1$, in which case ${\rm rank}(E_0 - \lambda_k I_{mT}) = m(T - 1)$.
	It suffices to show the argument for $m=1$, since the block case follows componentwise.
	
	There exists an eigenvector $r^k=
	[\lambda_k^{-1}\ \cdots\ \lambda_k^{-T}]^\top$
	satisfying $(E_0-\lambda_k I_T)r^k=0$ and $E_2^\top r^k=\lambda_k^{-T}=1$.
	To verify \eqref{eq:rank-ctrl-augsys}, let
	$[v^\top\ q^\top]$ belong to the left kernel of its matrix.
	Then
	\begin{align*}
		v^\top(\mathbf A_\iota-\lambda_k I_{n_\xi})&=0,\\
		v^\top\mathbf B_\iota E_2^\top
		+q^\top(E_0-\lambda_k I_T)&=0,\\
		v^\top\mathbf B_\iota+q^\top E_1&=0.
	\end{align*}
	Right multiplying the second equality by $r^k$ yields
	$v^\top\mathbf B_\iota=0$. Stabilizability of
	$(\mathbf A_\iota,\mathbf B_\iota)$ therefore implies $v=0$.
	The remaining equalities give
	$q^\top(E_0-\lambda_k I_T)=0$ and $q^\top E_1=0$, which imply
	$q=0$. Hence, \eqref{eq:rank-ctrl-augsys} holds for
	$\lambda=\lambda_k$.
	For
	$\lambda\notin\{\lambda_k:k=0,\ldots,T-1\}$,
	the matrix $E_0-\lambda I_{mT}$ is nonsingular. Moreover, noting that $I_m-E_2^\top(E_0-\lambda I_{mT})^{-1}E_1
	=
	\frac{\lambda^T}{\lambda^T-1}I_m$
	is nonsingular, block elimination reduces
	\eqref{eq:rank-ctrl-augsys} to the PBH rank condition for
	$(\mathbf A_\iota,\mathbf B_\iota)$. Therefore,
	$(A^a,B^a)$ is stabilizable.
	
	% There exists eigenvector $\zeta^k = [\lambda_k^{-T}\cdots \lambda_k^{-1}]^\top$ satisfying $\zeta^{k\top}(E_0 - \lambda_k I_{mT}) = 0$.
	% Further, we have $\zeta^{k\top} E_1 = e^{-{\rm i}2\pi kT/T} = 1$. 
	% % Furthermore, we have $E_2^\top \zeta^k = e^{-{\rm i}2\pi kT/T} = 1$.
	% The rank condition \eqref{eq:rank-ctrl-augsys} holds since $(A,B)$ is stabilizable.
	% For $\lambda \notin \{\lambda_k:k=0,\cdots,T-1\}$, the rank condition \eqref{eq:rank-ctrl-augsys} reduces to the stabilizability of $(A,B)$, and thus $(A^a,B^a)$ is stabilizable.
	
	The detectability of $(A^a,C^a)$ follows from \cite[Proposition 7]{kohler2021mpc-output-reg}; the argument is recalled for completeness.
	Detectability of $(A^a,C^a)$ is equivalent to 
	\begin{equation}\label{eq:rank-obsv-augsys}
		{\rm rank} \smat{\mathbf{A}_\iota - \lambda I_{n_\xi} & \mathbf{B}_\iota E^\top_2\\
			0 & E_0 - \lambda I_{mT}\\
			Z_1 & 0} = n_{\xi} + mT,
	\end{equation}
	for all $\lambda\in\C$ satisfying $|\lambda|\ge1$.
	As in the stabilizability argument, consider $\lambda_k = e^{{\rm i}2\pi k/T}$ for $k = 0,\cdots, T - 1$, in which case ${\rm rank}(E_0 - \lambda_k I_{mT}) = m(T - 1)$.
	Again, it suffices to prove the argument for $m=1$.
	There exists an eigenvector $\zeta^{k}_1 = [\lambda_k^{-1}\cdots \lambda_k^{-T}]^\top$ satisfying $(E_0 - \lambda_k I_{mT})\zeta^{k}_1 = 0$.
	Further, we have $E_2^\top \zeta^k_1 = e^{-{\rm i}2\pi kT/T} = 1$.
	Thus, the rank condition \eqref{eq:rank-obsv-augsys} is equivalent to the nonresonance condition \eqref{eq:nonresonance-cond} evaluated at $\lambda=\lambda_k$, which holds by Assumption~\ref{as:nonresonance-T}. Note that $\lambda_k$ need not be an eigenvalue of $S$; hence, Assumption~\ref{as:nonresonance-cond} alone is not sufficient at this step.
	For $\lambda \notin \{\lambda_k:k=0,\cdots,T-1\}$, the matrix $E_0-\lambda I_{mT}$ is nonsingular, and
	\eqref{eq:rank-obsv-augsys} reduces to the PBH rank condition for
	$(\mathbf A_\iota,Z_1)$. Hence, $(A^a,C^a)$ is detectable. 
\end{proof}

\section{Proof of Lemma \ref{lem:continue-prop-Y-noise}}
\label{app:proof-lem4}
\begin{proof}
	The proof has four steps. Step I bounds the one-step prediction error in
	terms of $\bar\varepsilon$. Step II constructs a shifted feasible candidate
	of length $L-1$ at time $t+1$. Step III bounds the cost of this candidate,
	and Step IV combines the resulting value-function estimate with the
	perturbed i-UIOSS inequality.
	
	\textbf{Step I.}
	Let $\hat\xi_{1|t}^*$ denote the one-step auxiliary state generated by
	the noise-free trajectory
	\begin{equation}\label{eq:relation-predict-real}
		\setlength{\abovedisplayskip}{3pt}
		\setlength{\belowdisplayskip}{3pt}
		H_{L+\iota}(y)g_t^*
		=\bar y_t^*+h_t^*-H_{L+\iota}(\varepsilon)g_t^*,
	\end{equation}
	with the optimal input and exosystem sequences at time $t$, and define
	$\delta_{1|t}^\xi:=\xi_{t+1}-\hat\xi_{1|t}^*$.
	
	From $Y_L^f(\tilde\varrho_t^a)\le\bar Y$ and nonnegativity of the
	objective terms in \eqref{eq:ddmpc-aug-noise:obj}, we get that
	\begin{equation}\label{eq:bound-g-h-t}
		\|h_t^*\|
		\le\sqrt{\frac{\bar Y\bar\varepsilon}{\lambda_h}},
		\qquad
		\|g_t^*-g_t^{\rm c}\|
		\le\sqrt{\frac{\bar Y}{\bar\varepsilon\lambda_g}}.
	\end{equation}
	Moreover, Proposition~\ref{prop:robust-center} and the bounds following
	\eqref{eq:hg=w-ss-noise} give $\|g_t^*\|
	\le
	\sqrt{\frac{\bar Y}{\bar\varepsilon\lambda_g}}
	+\sqrt{c_{\rm c}}
	+\sqrt{\frac{\bar J^s}{\lambda_{g^s}}}$.
	
	The input-history component and the exosystem prediction are exact.
	Hence, by shifting the output-history component in
	\eqref{eq:relation-predict-real}, the one-step prediction error consists
	only of the online measurement error, the noisy-Hankel term
	$H_{L+\iota}(\varepsilon)g_t^*$, and the slack $h_t^*$.
	Each relevant row block of $H_{L+\iota}(\varepsilon)$ is bounded by a
	constant multiple of $\bar\varepsilon$. Together with
	\eqref{eq:bound-g-h-t} and the auxiliary output equation, this yields
	functions $b_y,b_\xi\in\mathcal K_\infty$ such that
	\begin{equation}\label{eq:output-pred-error-yt-y0t-norm}
		\|y_t-\bar y_{0|t}^*\|
		\le b_y(\bar\varepsilon),
		\qquad
		\|\delta_{1|t}^\xi\|
		\le b_\xi(\bar\varepsilon).
	\end{equation}
	
	Since the input memory is generated from the applied inputs, its
	one-step prediction is exact. Thus, with
	$\hat\xi_{1|t}^{a,*}
	:=\smat{\hat\xi_{1|t}^*\\\bar\zeta_{1|t}^*}$, we get that $
	\|\delta_{1|t}^{\xi^a}\|
	\le b_\xi(\bar\varepsilon)$, where $\delta_{1|t}^{\xi^a}
	:=\xi_{t+1}^a-\hat\xi_{1|t}^{a,*}
	=\smat{\delta_{1|t}^{\xi}\\0_{mT\times1}}$.

	\textbf{Step II.}
	We construct a candidate of length $L-1$ at time $t+1$ by shifting the
	previously optimal physical-input sequence:
	\begin{equation}\label{eq:baru-t+1}
		\bar u_{[-\iota,L-2]|t+1}
		:=\bar u^*_{[-\iota+1,L-1]|t},
	\end{equation}
	and set
	\begin{subequations}
		\label{eq:shift-aug-candidate}
		\begin{align}
			\bar\zeta_{k|t+1}
			&:=\bar\zeta^*_{k+1|t},
			&&k=0,\ldots,L-1,\\
			\bar u^a_{k|t+1}
			&:=\bar u^{a,*}_{k+1|t},
			&&k=0,\ldots,L-2.\label{eq:shift-aug-candidate:u}
		\end{align}
	\end{subequations}
	Since the first optimal physical input is applied at time $t$, the
	shifted memory state satisfies the required initialization. The augmented
	dynamics and the physical-input constraint follow directly from the
	previous feasible solution; in particular,
	$\bar u_{k|t+1}=\bar u^*_{k+1|t}\in\mathbb U$ for
	$k=0,\ldots,L-2$.
	
	It remains to construct the corresponding Hankel coefficient and slack.
	Let $	H_{xum}:=
	\smat{
		H_1(x_{[0,N-L-\iota+1]})\\
		H_{L+\iota-1}(u)\\
		H_1(\mathcal M_{[0,N-L-\iota+1]})
	}$.
%	\[
%	H_{xum}:=
%	\smat{
%		H_1(x_{[0,N-L-\iota+1]})\\
%		H_{L+\iota-1}(u)\\
%		H_1(\mathcal M_{[0,N-L-\iota+1]})
%	}.
%	\]
	The assumed persistence of excitation of order $L+\iota$ implies that
	$H_{xum}$ has full row rank at the shortened depth. Moreover,
	observability of \eqref{eq:sys} uniquely determines
	$x_{t+1-\iota}$ from the last $\iota$ noise-free I/O samples. Hence, a
	right inverse of $H_{xum}$ provides
	\begin{equation}\label{eq:barg-t+1}
		\bar g_{t+1}
		=H_{xum}^{\dag}
		\smat{
			x_{t+1-\iota}\\
			\bar u_{[-\iota,L-2]|t+1}\\
			\bar M_{t+1-\iota}
		}.
	\end{equation}
	Choose the output and slack candidates as
	\begin{align}
		\bar y_{[-\iota,L-2]|t+1}
		&=H_{L+\iota-1}(y)\bar g_{t+1}
		+\smat{
			\varepsilon_{[t+1-\iota,t]}\\
			0_{p(L-1)\times1}
		},
		\label{eq:nominal-output-t+1}\\
		\bar h_{[-\iota,L-2]|t+1}
		&=H_{L+\iota-1}(\varepsilon)\bar g_{t+1}
		-\smat{
			\varepsilon_{[t+1-\iota,t]}\\
			0_{p(L-1)\times1}
		}.
		\label{eq:cand-barh-t1}
	\end{align}
	These choices satisfy the noisy Hankel equation and the measured
	initial-condition constraint by construction. Together with
	\eqref{eq:baru-t+1}--\eqref{eq:shift-aug-candidate}, this proves
	feasibility of the shortened candidate.
	
	\textbf{Step III.}
	We first compare the shifted output with the previous optimal output.
	Since the two predictions use the same shifted input and exosystem
	sequences, the auxiliary dynamics \eqref{eq:sys:aux} give
	\begin{align*}
		\delta^y_{k|t+1}
		&:=\bar y_{k|t+1}-\bar y^*_{k+1|t}\\
		&=Z_1\mathbf A_\iota^k\delta_{1|t}^\xi
		+h^*_{k+1|t}
		-\bigl(H_{L+\iota}(\varepsilon)g_t^*\bigr)_{k+1}.
	\end{align*}
	Combining this identity with \eqref{eq:bound-g-h-t} and
	\eqref{eq:output-pred-error-yt-y0t-norm}, and using the fact that the
	horizon is fixed, gives some $b_\delta\in\mathcal K_\infty$ such that
	\begin{equation}\label{eq:delta-y-square-up-bound}
		\sum_{k=0}^{L-2}
		\|\delta^y_{k|t+1}\|_Q^2
		\le b_\delta(\bar\varepsilon).
	\end{equation}
	
	It remains to bound the coefficient and slack regularization terms.
	Partition $\Pi^a=\smat{\Pi_\xi\\\Pi_\zeta^a}$. The regulator equations
	and \eqref{eq:sys:aug-xi} imply $\bar u_{k|t}-\Gamma_\xi\bar\omega_{k|t}
	=
	E_2^\top
	(\bar\zeta_{k|t}-\Pi_\zeta^a\bar\omega_{k|t})
	+\bar u^a_{k|t}$.
	On the sublevel set $Y_L^f(\tilde\varrho_t^a)\le\bar Y$, the lower
	bound on $W^a$ and the objective in
	\eqref{eq:ddmpc-aug-noise:obj} uniformly bound the initial augmented
	error and the optimal augmented-input sequence. Together with the
	one-step error estimate from Step I, the finite-horizon augmented
	dynamics, and the observability reconstruction of
	$x_{t+1-\iota}$, this uniformly bounds the state and physical-input
	quantities entering the right-inverse construction
	\eqref{eq:barg-t+1}. Hence, Proposition~\ref{prop:robust-center} and
	$\|g_{t+1}^{s*}-\tilde g_{t+1}^s\|^2\le c_{g^s}$ give a constant
	$c_g>0$, independent of $\bar\varepsilon$, such that
	\begin{equation}\label{eq:bound-candidate-g-center}
		\|\bar g_{t+1}-g_{t+1}^{\rm c}\|^2
		\le c_g^2.
	\end{equation}
	The same estimates and the construction
	\eqref{eq:cand-barh-t1} give a constant $b_h>0$, such that $\|\bar h_{t+1}\|
	\le\bar\varepsilon b_h$.
%	\begin{equation}\label{eq:up-bound-barh-t1}
%		\|\bar h_{t+1}\|
%		\le\bar\varepsilon b_h.
%	\end{equation}
	
	Define $\ell_{0|t}^{a,*}
	:=
	\|\bar y_{0|t}^*\|_Q^2
	+\|\bar u_{0|t}^{a,*}\|_R^2$.
	Using candidate feasibility, $\bar u^a$ in \eqref{eq:shift-aug-candidate:u}, \eqref{eq:delta-y-square-up-bound}, and
	$J_L^{f,*}\le\bar Y$, we obtain
	\begin{align}
		J_{L-1}^{f,*}(\tilde\varrho_{t+1}^a)
		&\le
		J_L^{f,*}(\tilde\varrho_t^a)
		-\ell_{0|t}^{a,*}\nonumber\\
		&\quad
		+2\sqrt{\bar Y b_\delta(\bar\varepsilon)}
		+b_\delta(\bar\varepsilon)
		+\lambda_g c_g^2\bar\varepsilon
		+\lambda_h b_h^2\bar\varepsilon\nonumber\\
		&=:
		J_L^{f,*}(\tilde\varrho_t^a)
		-\ell_{0|t}^{a,*}
		+\alpha_1(\bar\varepsilon),
		\label{eq:decrease-J-noise}
	\end{align}
	where $\alpha_1\in\mathcal K_\infty$.
	
	\textbf{Step IV.}
	The applied augmented input satisfies
	$u_t^a=\bar u_{0|t}^{a,*}$. Moreover,
	\eqref{eq:output-pred-error-yt-y0t-norm} and the sublevel bound imply the
	existence of $\alpha_y\in\mathcal K_\infty$ such that
	\begin{equation}\label{eq:actual-predicted-stage-cost}
		\|y_t\|_Q^2
		\le\|\bar y_{0|t}^*\|_Q^2+\alpha_y(\bar\varepsilon).
	\end{equation}
	Applying \eqref{eq:iuioss-perturbed-W} and
	\eqref{eq:actual-predicted-stage-cost} gives
	\begin{align}\label{eq:decrease-W-noise}
		W^a(\tilde\varrho_{t+1}^a)
		&\le W^a(\tilde\varrho_t^a)
		+\|\bar u_{0|t}^{a,*}\|_R^2
		+\|\bar y_{0|t}^*\|_Q^2\nonumber\\
		&\quad-\epsilon_o\sigma^a(\tilde\varrho_t^a)
		+\alpha_W(\bar\varepsilon)+\alpha_y(\bar\varepsilon).
	\end{align}
	Adding \eqref{eq:decrease-J-noise} and
	\eqref{eq:decrease-W-noise} proves
	\eqref{eq:continue-prop-Y-noise} with
	$\alpha_3:=\alpha_1+\alpha_W+\alpha_y\in\mathcal K_\infty$.
\end{proof}

\section{Proof of Theorem \ref{thm:ddmpc-stab-noise}}
\label{app:proof-thm3}
\begin{proof}
	The proof follows the same finite-horizon argument as Theorem
	\ref{thm:stab-ori-ddmpc}; hence, we only indicate the modifications caused
	by the perturbation terms in Lemma \ref{lem:continue-prop-Y-noise}.
	
	\textbf{Step I.}
	The lower bound in \eqref{eq:up-low-bound-Y-noise} follows directly from
	$J_L^{f,*}\ge0$ and Proposition~\ref{prop:i-UIOSS}. For the upper bound,
	Proposition~\ref{prop:aug-ctrl-detect-equ-ori-ctrl-detect} and Assumption
	\ref{as:strict-reg-input} give a local feedback
	$u^{a,\mathrm{loc}}=K^a(\xi^{a,\mathrm{loc}}-\Pi^a\omega)$ and constants
	$\delta_{\rm loc},\gamma_s>0$ such that, whenever
	$\sigma^a(\varrho_t^a)\le\delta_{\rm loc}$, the resulting physical inputs belong to
	$\mathbb U$ and $\sum_{j=0}^{\infty}
	\ell^a(\varrho_{j|t}^{a,\mathrm{loc}},
	u_{j|t}^{a,\mathrm{loc}})
	\le\gamma_s\sigma^a(\varrho_t^a)$.
	This is the augmented counterpart of the local construction in the nominal
	proof.
	
	Since
	$\|\tilde\xi_t^a-\xi_t^a\|\le\sqrt{p\iota}\bar\varepsilon$, we have $\sigma^a(\varrho_t^a)
	\le
	2\sigma^a(\tilde\varrho_t^a)
	+2p\iota\bar\varepsilon^2$.
	Hence, by choosing $\delta_a<\delta_{\rm loc}/2$ and the admissible noise
	bound sufficiently small, $\sigma^a(\tilde\varrho_t^a)\le\delta_a$
	implies $\sigma^a(\varrho_t^a)\le\delta_{\rm loc}$, so the same local
	construction remains admissible.
	The trajectory parametrization in Theorem~\ref{thm:fund-lem-exo},
	together with the right-inverse and noise-compensating slack construction
	in the proof of Lemma~\ref{lem:continue-prop-Y-noise}, provides a feasible
	robust DDPC candidate for every finite segment of this local trajectory.
	Moreover, the preceding bound on $\sigma^a(\varrho_t^a)$ and
	Proposition~\ref{prop:robust-center} imply that the state perturbation and the
	additional coefficient and slack regularization terms can be collected into
	a function in $\mathcal K_\infty$.
	Therefore, for some $\gamma_{\mathrm{loc}}>0$ and
	$\alpha_L\in\mathcal K_\infty$, it follows that $	Y_L^f(\tilde\varrho_t^a)
	\le
	\gamma_{\mathrm{loc}}\sigma^a(\tilde\varrho_t^a)
	+\alpha_L(\bar\varepsilon)$, where $\sigma^a(\tilde\varrho_t^a)\le\delta_a$.
	Combining this with $Y_L^f\le\bar Y$ for
	$\sigma^a(\tilde\varrho_t^a)>\delta_a$ proves \eqref{eq:up-low-bound-Y-noise} with
	$\gamma_{\bar Y}:=\max\{\gamma_{\mathrm{loc}},\bar Y/\delta_a\}$.
	
	\textbf{Step II.}
	Let the superscript $\mathrm c$ denote the shortened candidate constructed
	in the proof of Lemma~\ref{lem:continue-prop-Y-noise}, and set
	$Y_{L-1}^{f,\mathrm c}:=J_{L-1}^{f,\mathrm c}
	+W^a(\tilde\varrho_{t+1}^a)$. The estimates used there hold for this
	candidate before minimization; thus,
	\begin{equation}\label{eq:short-candidate-Yf}
		Y_{L-1}^{f,\mathrm c}(\tilde\varrho_{t+1}^a)
		\le Y_L^f(\tilde\varrho_t^a)
		-\epsilon_o\sigma^a(\tilde\varrho_t^a)
		+\alpha_3(\bar\varepsilon).
	\end{equation}
	
	Let $\{\bar\varrho_{k|t+1}^{a,\mathrm c}\}_{k=0}^{L-1}$ denote the
	nominal augmented trajectory associated with this candidate.
	Applying \eqref{eq:iuioss-decrease-W} along this trajectory and summing
	from $k=0$ to $L-2$ gives
	\begin{equation}\label{eq:sum-sigma-short}
		\epsilon_o\sum_{k=0}^{L-2}
		\sigma^a(\bar\varrho_{k|t+1}^{a,\mathrm c})
		\le W^a(\varrho_{t+1}^a)
		+\sum_{k=0}^{L-2}
		\ell^a(\bar\varrho_{k|t+1}^{a,\mathrm c},
		\bar u_{k|t+1}^{a,\mathrm c}),
	\end{equation}
	where the nonnegative terminal storage has been dropped.
	
	On the sublevel set under consideration, quadratic continuity of $W^a$ and
	$\|\tilde\xi_{t+1}^a-\xi_{t+1}^a\|
	\le\sqrt{p\iota}\bar\varepsilon$ imply $W^a(\varrho_{t+1}^a)
	\le W^a(\tilde\varrho_{t+1}^a)+\alpha_5(\bar\varepsilon)$
	for some $\alpha_5\in\mathcal K_\infty$.
	Since the regularization terms are nonnegative, \eqref{eq:short-candidate-Yf}, 
	\eqref{eq:sum-sigma-short}, and the upper
	bound established in Step I yield
	\setlength{\abovedisplayskip}{3pt}
	\setlength{\belowdisplayskip}{3pt}
	\begin{align}
		\epsilon_o\sum_{k=0}^{L-2}
		\sigma^a(\bar\varrho_{k|t+1}^{a,\mathrm c})
		&\le c_{\bar Y}\sigma^a(\tilde\varrho_t^a)
		+\alpha_6(\bar\varepsilon),
		\label{eq:sum-sigma-short-bound}
	\end{align}
	where
	$c_{\bar Y}:=\max\{\gamma_{\bar Y}-\epsilon_o,0\}$ and
	$\alpha_6:=\alpha_L+\alpha_3+\alpha_5$.
	
	Hence, by the same averaging argument as in Step II of the proof of
	Theorem~\ref{thm:stab-ori-ddmpc}, there exists
	$k_x\in\{0,\ldots,L-2\}$ such that
	\begin{equation}\label{eq:kx-aug}
		\sigma^a(\bar\varrho_{k_x|t+1}^{a,\mathrm c})
		\le\frac{c_{\bar Y}}{\epsilon_o(L-1)}
		\sigma^a(\tilde\varrho_t^a)
		+\frac{\alpha_6(\bar\varepsilon)}{\epsilon_o(L-1)}.
	\end{equation}
	The same argument using only $Y_L^f(\tilde\varrho_t^a)\le\bar Y$ gives $\sigma^a(\bar\varrho_{k_x|t+1}^{a,\mathrm c})
	\le\frac{\bar Y+\alpha_3(\bar\varepsilon)
		+\alpha_5(\bar\varepsilon)}{\epsilon_o(L-1)}$.
	Therefore, if $L>1+\frac{2\bar Y}{\epsilon_o\delta_a}$
	and $\bar\varepsilon$ is sufficiently small such that
	$\alpha_3(\bar\varepsilon)+\alpha_5(\bar\varepsilon)\le\bar Y$, then
	$\sigma^a(\bar\varrho_{k_x|t+1}^{a,\mathrm c})\le\delta_a$.
	The local controller from Step I can thus be appended at $k_x$ while
	preserving the physical-input constraint.
	
	\textbf{Step III.}
	Append the local feedback at $k_x$ to the first $k_x$ steps of the shortened
	candidate and propagate \eqref{eq:sys:aug}. This gives a length-$L$
	augmented trajectory whose physical inputs lie in $\mathbb U$.
	Persistence of excitation provides a coefficient $\bar g_{t+1}^f$ for this
	concatenated nominal trajectory at depth $L+\iota$. Choose
$\bar y_{[-\iota,L-1]|t+1}^f
=H_{L+\iota}(y)\bar g_{t+1}^f
+\smat{\varepsilon_{[t+1-\iota,t]}\\0_{pL\times1}}$ and $\bar h_{t+1}^f
=H_{L+\iota}(\varepsilon)\bar g_{t+1}^f
-\smat{\varepsilon_{[t+1-\iota,t]}\\0_{pL\times1}}$.
%	\begin{align*}
%		\bar y_{[-\iota,L-1]|t+1}^f
%		&=H_{L+\iota}(y)\bar g_{t+1}^f
%		+\smat{\varepsilon_{[t+1-\iota,t]}\\0_{pL\times1}},\\
%		\bar h_{t+1}^f
%		&=H_{L+\iota}(\varepsilon)\bar g_{t+1}^f
%		-\smat{\varepsilon_{[t+1-\iota,t]}\\0_{pL\times1}}.
%	\end{align*}
	Then constraints \eqref{eq:ddmpc-aug-noise:hankel} and
	\eqref{eq:ddmpc-aug-noise:init} hold.
	
	The same right-inverse estimates as in the proof of
	Lemma~\ref{lem:continue-prop-Y-noise}, now applied at depth $L+\iota$,
	together with Proposition~\ref{prop:robust-center}, which controls
	$g_{t+1}^{\rm c}-\tilde g_{t+1}^s$, give constants $c_g^f,c_h^f>0$
	independent of $\bar\varepsilon$ such that
	\begin{equation}
		\|\bar g^f_{t+1}-g_{t+1}^{\rm c}\|\le c_g^f,\qquad
		\|\bar h_{t+1}^f\|\le\bar\varepsilon c_h^f.
	\end{equation}
	
	The augmented input is shifted exactly, and the output mismatch over the
	prefix is bounded as in the proof of
	Lemma~\ref{lem:continue-prop-Y-noise}, whereas the local-tail cost is bounded
	by $\gamma_s$ times the estimate in Step II. Together with the
	regularization bounds and optimality, this yields some
	$\alpha_J\in\mathcal K_\infty$ satisfying
	\begin{align}
		J_L^{f,*}(\tilde\varrho_{t+1}^a)
		&\le J_L^{f,*}(\tilde\varrho_t^a)-\ell_{0|t}^{a,*}\nonumber\\
		&\quad+\frac{\gamma_s c_{\bar Y}}{\epsilon_o(L-1)}
		\sigma^a(\tilde\varrho_t^a)+\alpha_J(\bar\varepsilon).
	\end{align}
	Here $\alpha_J$ collects the prefix perturbation, the term
	$\gamma_s\alpha_6(\bar\varepsilon)/[\epsilon_o(L-1)]$, and the
	coefficient and slack penalties
	$\lambda_g\bar\varepsilon(c_g^f)^2$ and
	$\lambda_h\bar\varepsilon(c_h^f)^2$.
	
	Adding the perturbed storage estimate \eqref{eq:decrease-W-noise} proves
	\eqref{eq:decrease-Y-noise} with $	\tilde\alpha_L:=\epsilon_o-
	\frac{\gamma_s c_{\bar Y}}{\epsilon_o(L-1)}>0$ and $\alpha_8:=\alpha_J+\alpha_W+\alpha_y\in\mathcal K_\infty$,
	where positivity holds for sufficiently large $L$.
	
	It remains to account for the additive perturbations when proving
	invariance. Choose $L>\tilde L_{\bar Y}$ large enough for both the local-entry
	condition in Step II and $\tilde\alpha_L>0$, and then choose
	$\hat\varepsilon_L>0$ such that, for
	$\bar\varepsilon\le\hat\varepsilon_L$, $\alpha_L(\bar\varepsilon)+\alpha_8(\bar\varepsilon)<\bar Y$ and $\alpha_8(\bar\varepsilon)\le
	\frac{\tilde\alpha_L}{\gamma_{\bar Y}}
	\bigl(\bar Y-\alpha_L(\bar\varepsilon)
	-\alpha_8(\bar\varepsilon)\bigr)$.
	If $Y_L^f(\tilde\varrho_t^a)\le
	\bar Y-\alpha_8(\bar\varepsilon)$, invariance follows directly from
	\eqref{eq:decrease-Y-noise}. Otherwise, the upper bound in
	\eqref{eq:up-low-bound-Y-noise} gives $\sigma^a(\tilde\varrho_t^a)
	\ge
	\frac{\bar Y-\alpha_8(\bar\varepsilon)
		-\alpha_L(\bar\varepsilon)}{\gamma_{\bar Y}}$.
	Hence, by the chosen noise bound, the decrease term dominates
	$\alpha_8(\bar\varepsilon)$ and
	$Y_L^f(\tilde\varrho_{t+1}^a)\le
	Y_L^f(\tilde\varrho_t^a)\le\bar Y$.
	Thus, $Y_L^f\le\bar Y$ is forward invariant, and the candidate above proves
	recursive feasibility and $u_t\in\mathbb U$ by induction.
	
	Finally, \eqref{eq:up-low-bound-decrease-Y-noise} gives practical stability
	with an ultimate neighborhood that vanishes as
	$\bar\varepsilon\to0$, and
	$\|\tilde\xi_t-\Pi_\xi\omega_t\|^2\le
	\sigma^a(\tilde\varrho_t^a)$ transfers this property to the original
	regulation error.
	Lemma~\ref{lem:aug-mpc-equ-robust-mpc} transfers these conclusions from
	\eqref{eq:ddmpc-aug-noise} to \eqref{eq:ddmpc-incre-noise}, which completes the proof.
\end{proof}

\end{document}